\documentclass[]{aa}  

\usepackage{graphicx}
\usepackage{txfonts}
\usepackage{adjustbox}
\usepackage{caption}
\usepackage{float}
\usepackage[flushleft]{threeparttable}
\usepackage{pdfpages}
\usepackage{longtable}
\usepackage{textcomp}
\usepackage{multirow}
\usepackage{enumerate}
\usepackage{hhline}
\usepackage{comment}
\usepackage{rotating}
\usepackage{array}
\usepackage{hyperref} 
\begin{document}

   \title{The deep \textit{Chandra} survey in the SDSS J1030+0524 field}

   %\subtitle{}

   \author{R. Nanni
          \inst{1,2,3}
          \and R. Gilli
          \inst{1}
          \and C. Vignali
          \inst{2,1}
          \and M. Mignoli
          \inst{1}
          \and A. Peca
          \inst{1,4}
          \and S. Marchesi
          \inst{1,5}
          \and M. Annunziatella
           \inst{6}
          \and M. Brusa
          \inst{2}
          \and F. Calura
          \inst{1}
	 \and N. Cappelluti
          \inst{4,7,8}
          \and M. Chiaberge
          \inst{9,10}
          \and A. Comastri
          \inst{1}
          \and K. Iwasawa
          \inst{11,12}
          \and G. Lanzuisi
          \inst{1}
          \and E. Liuzzo
          \inst{13}
          \and D. Marchesini
          \inst{6}
          \and I. Prandoni
          \inst{13}
          \and P. Tozzi
          \inst{14}
          \and F. Vito
          \inst{15,16}
          \and G. Zamorani
          \inst{1}
          \and C. Norman
          \inst{9,10}
          }

   \institute{INAF - Osservatorio di Astrofisica e Scienza dello Spazio di Bologna, via Gobetti 93/3 - 40129 Bologna - Italy
         \and
             Dipartimento di Fisica e Astronomia, Universit\`a degli Studi di Bologna, via Gobetti 93/2, 40129 Bologna, Italy
          \and
          Department of Physics, University of California, Santa Barbara, CA 93106-9530, USA
          \and
           Physics Department, University of Miami, Coral Gables, FL 33124
           \and
           Department of Physics and Astronomy, Clemson University,  Kinard Lab of Physics, Clemson, SC 29634, USA
           \and
           Physics Department, Tufts University, 574 Boston Avenue, Medford, 02155, MA
             \and
             Yale Center for Astronomy and Astrophysics, P.O. Box 208121, New Haven, CT 06520, USA 
             \and
             Department of Physics, Yale University, P.O. Box 208121, New Haven, CT 06520, USA 
             \and
             Space Telescope Science Institute, 3700 San Martin Dr., Baltimore, MD 21210, USA
             \and
             Johns Hopkins University, 3400 N. Charles Street, Baltimore, MD 21218, USA
             \and
             Institut de Ci\`encies del Cosmos (ICCUB), Universitat de Barcelona (IEEC-UB), Mart\'i i Franqu\`es, 1, 08028 Barcelona, Spain
             \and
	5 ICREA, Pg. Llu\'is Companys 23, 08010 Barcelona, Spain
             \and
             INAF - Istituto di Radioastronomia, via Gobetti 101, I-40129 Bologna, Italy
             \and
             INAF, Osservatorio Astrofisico di Arcetri, Largo E. Fermi 5, I-50125 Firenze, Italy
             \and
             Instituto de Astrofisica and Centro de Astroingenieria, Facultad de Fisica, Pontificia Universidad Catolica de Chile, Casilla 306, Santiago 22, Chile
             \and
             Chinese Academy of Sciences South America Center for Astronomy, National Astronomical Observatories, CAS, Beijing 100012, PR China
              \\             }

   \date{}
 
  \abstract
  {We present the X-ray source catalog for the $\sim$479 ks \textit{Chandra} exposure of the SDSS J1030+0524 field, that is centered on a region that shows the best
evidence to date of an overdensity around a $z>6$ quasar, and also includes a galaxy overdensity around a Compton-thick Fanaroff-Riley type II (FRII) radio galaxy at $z = 1.7$.
  %This is currently the fifth deepest extragalactic X-ray survey, and the field is part of the Multiwavelength Survey by Yale-Chile (MUSYC).
  Using \textit{wavdetect} for initial source detection and ACIS Extract for source photometry and significance assessment, we create preliminary catalogs of sources that are detected in the full (0.5-7.0 keV), soft (0.5-2.0 keV), and hard (2-7 keV) bands, respectively. We produce X-ray simulations that mirror our \textit{Chandra} observation to filter our preliminary catalogs and get a completeness level of $>91\%$ and a reliability level of $\sim95\%$ in each band. The catalogs in the three bands are then matched into a final main catalog of 256 unique sources. Among them, 244, 193, and 208 are detected in the full, soft, and hard bands, respectively. 
The \textit{Chandra} observation covers a total area of 335 arcmin$^2$, and reaches flux limits over the central few square arcmins of $\sim3\times10^{-16}$, $6\times10^{-17}$, and $2\times10^{-16}$ erg cm$^{-2}$ s$^{-1}$ in the full, soft, and hard bands, respectively
This makes J1030 field the fifth deepest extragalactic X-ray survey to date. The field is part of the Multiwavelength Survey by Yale-Chile (MUSYC), and is also covered by optical imaging data from the Large Binocular Camera (LBC) at the Large Binocular Telescope (LBT), near-IR imaging data from the Canada France Hawaii Telescope WIRCam (CFHT/WIRCam), and \textit{Spitzer} IRAC.
Thanks to its dense multi-wavelength coverage, J1030 represents a legacy field for the study of large-scale structures around distant accreting supermassive black holes.
Using a likelihood ratio analysis, we associate multi-band ($r$, $z$, $J$, and $4.5\, \mu m$) counterparts for 252 (98.4\%) of the 256 \textit{Chandra} sources, with an estimated reliability of 95\%. Finally, we compute the cumulative number of sources in each X-ray band, finding that they are in general agreement with the results from the \textit{Chandra} Deep Fields.
  }
  
               
   \maketitle

%
%-------------------------------------------------------------------

\section{Introduction}\label{intro}

Deep X-ray surveys provide a highly efficient method to pinpoint growing black holes in active galactic nuclei (AGN) across a wide range of redshifts, and offer insights about the demographics, physical properties, and interactions with the environment of super massive black holes (SMBHs).
Furthermore, they are primary tools to study the diffuse emission of clusters and groups, as well as X-ray binaries in distant star-forming galaxies: the \textit{Chandra} Deep Field-South (CDF-S; \citealt{Luo17}), the \textit{Chandra} Deep Field-North (CDF-N; \citealt{Xue16}), the AEGIS-X survey (\citealt{Nandra15}), the \textit{Chandra} UKIDSS Ultra Deep Survey (X-UDS; \citealt{Kocevski18}), and the COSMOS Legacy survey (\citealt{Civ16}; \citealt{Mar16}) are at present some of the main surveys to investigate the deep X-ray Universe.

While shallow large area surveys are essential to cover large portions of the sky, avoiding field-to-field variance problems and providing a global view of the most luminous X-ray sources (e.g., XMM-XXL and Stripe 82X surveys; \citealt{Menzel16}; \citealt{LaMassa16}), deep X-ray surveys are capable of reaching extremely faint flux levels and thus earlier cosmic epochs.
In addition, at a given redshift, deep surveys can probe objects with intrinsically low X-ray luminosities (that are generally more representative of the source population) including star-forming galaxies (this population is dominant at fluxes $f\le 10^{-17}$ erg s$^{-1}$ cm$^{-2}$ in the 0.5-2 keV band; \citealt{Leh12}), as well as intrinsically luminous sources that are dimmed by strong nuclear obscuration (e.g., \citealt{Norman04}; \citealt{Comastri11}; \citealt{Gilli11}).

\begin{figure*}
 \centering
\includegraphics[height=9cm, width=9cm, keepaspectratio]{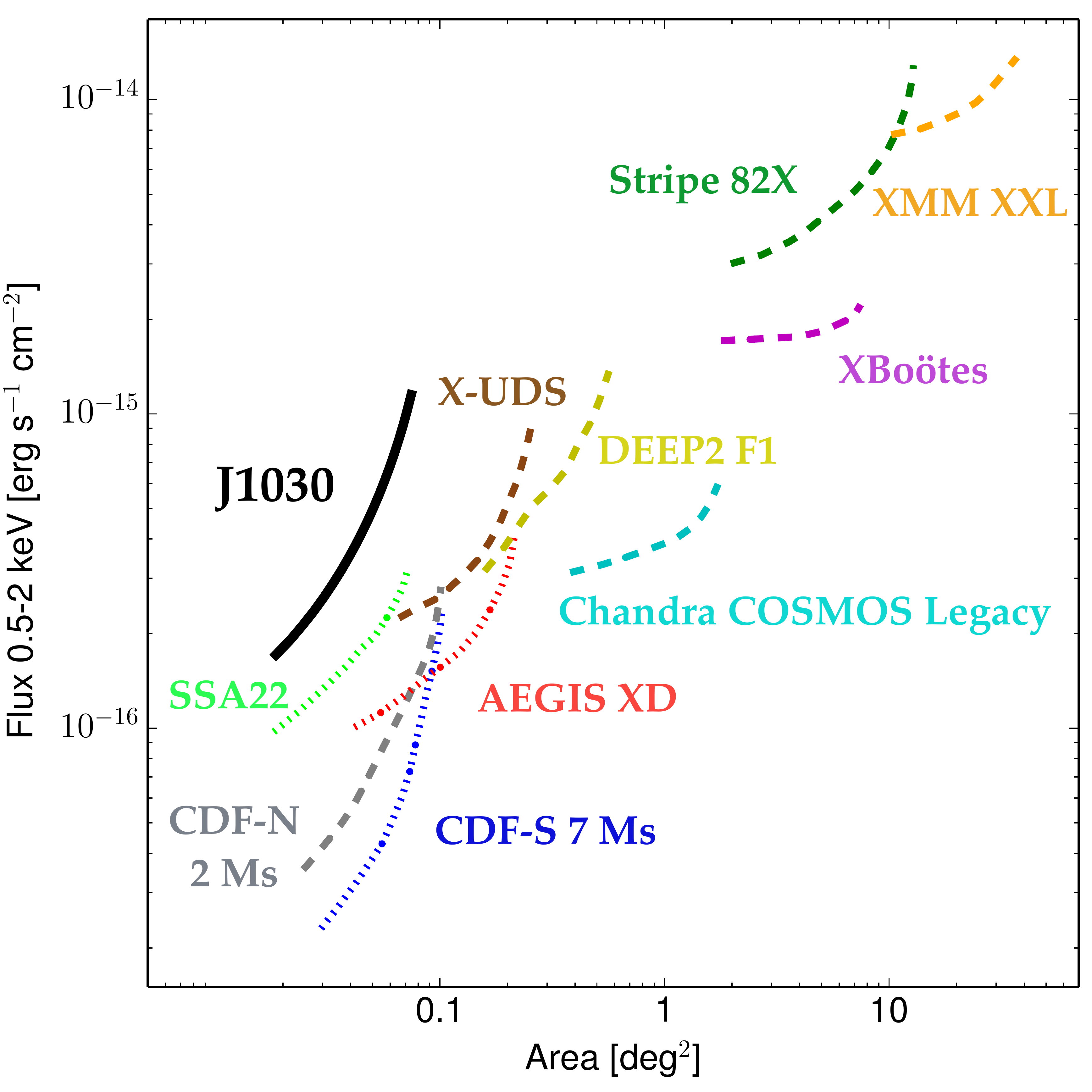}\includegraphics[height=9cm, width=9cm, keepaspectratio]{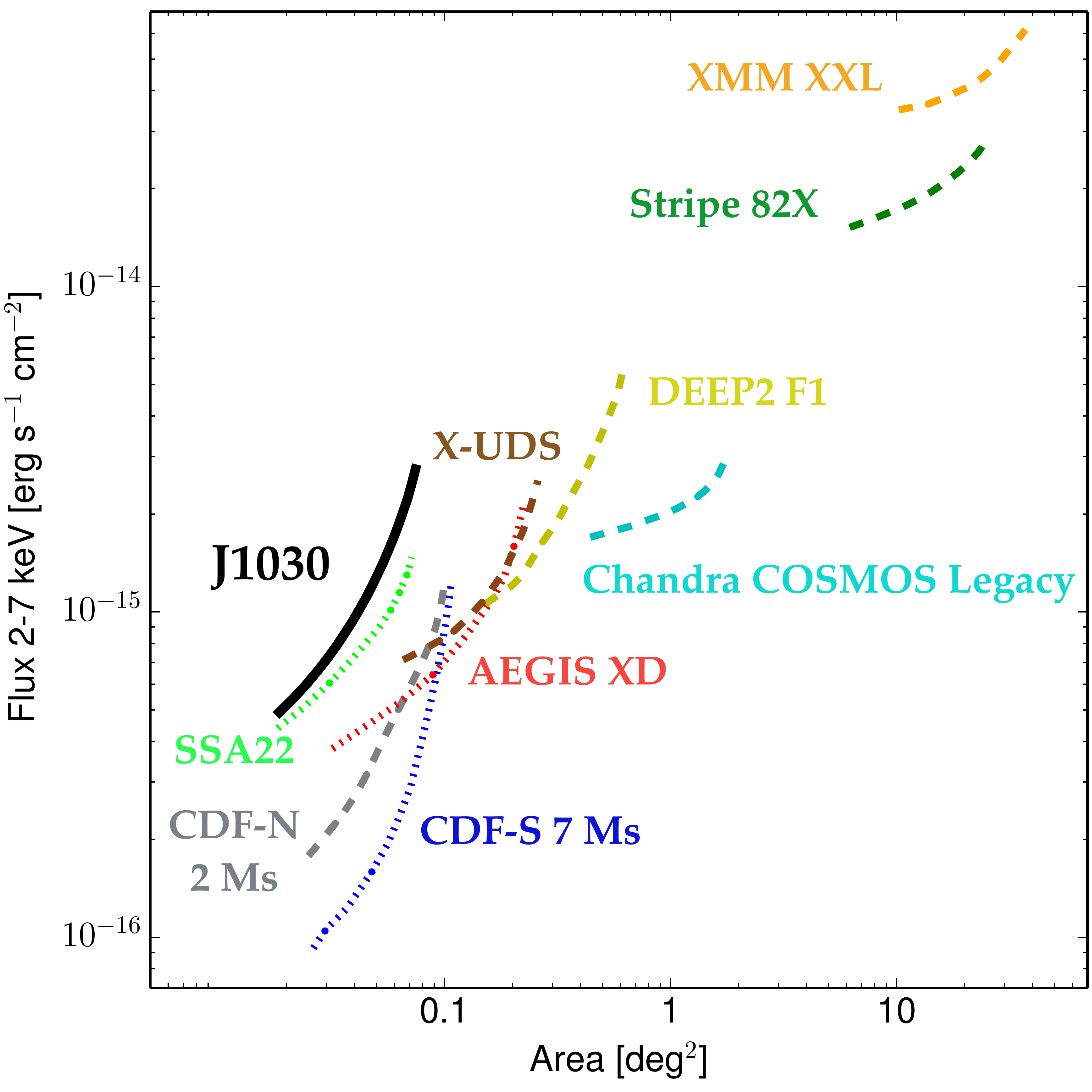}
\caption{Area-flux curves for different deep and moderately deep \textit{Chandra} surveys in the soft (left panel) and hard (right panel) bands. Each survey has been plotted using each sensitivity curve starting from the flux corresponding to 80\% of the maximum area for that survey to the flux corresponding to 20\% of the total area. The reported surveys are from: this work (black line),  \citet{Luo17} (blue dotted line), \citet{Xue16} (gray dashed line), \citet{Nandra15} (red dotted line), \citet{Lehmer09} (light green dotted line), \citet{Kocevski18} (brown dashed line), \citet{Civ16} (cyan dashed line), \citet{LaMassa16} (dark green dotted line), \citet{Menzel16} (orange dashed line), \citet{Goulding12} (yellow dashed line), and \citet{Murray05} (magenta dashed line). Despite the shorter exposure ($\sim$400 ks), the SSA22 survey is deeper in the soft band than the J1030 one due to the \textit{Chandra} effective area degradation (equal to $\sim25\%$ at 1.4 keV) in this band.}
\label{fig:surveys}
\end{figure*}

So far, the deepest four X-ray surveys are: the CDF-S, with an exposure of $\sim$7 Ms over an area of 484.2 arcmin$^2$ (\citealt{Luo17}), the CDF-N, with an exposure of $\sim$2 Ms over an area of 447.5 arcmin$^2$ (\citealt{Xue16}), the AEGIS-X survey, with an exposure of $\sim$800 ks over an area of $\sim1040$ arcmin$^2$ (\citealt{Nandra15}), and the SSA22 survey, with an exposure of $\sim$400 ks over an area of $\sim330$ arcmin$^2$ (\citealt{Lehmer09}). These surveys achieved unprecedented X-ray sensitivity with flux limits in their inner square arcmins of $\sim1.9,0.6,2.7 \times 10^{-17}$ erg s$^{-1}$ cm$^{-2}$ for CDF-S, $\sim3.5,1.2,5.9 \times 10^{-17}$ erg s$^{-1}$ cm$^{-2}$ for CDF-N, $\sim1.5,3.4,2.5 \times 10^{-16}$ erg s$^{-1}$ cm$^{-2}$ for AEGIS-X, and $\sim1.7,0.6,3.0 \times 10^{-16}$ erg s$^{-1}$ cm$^{-2}$ for SSA22, in the full (0.5-7 keV), soft (0.5-2 keV), and hard (2-7 keV) bands, respectively. In Fig. \ref{fig:surveys} we show the area-flux curves for the deepest \textit{Chandra} surveys achieved so far, including the flux limits computed for J1030+0524 (hereafter J1030) in \S \ref{logNlogS}.

In this paper, we present the point-source catalog derived from the $\sim$479 ks \textit{Chandra} exposure of the J1030 field that we obtained in 2017. This X-ray field has a nominal aim point centered on the quasar (QSO) SDSS J1030+0525 at $z = 6.31$ (\citealt{Fan01}). This QSO was one of the first $z\sim6$ QSOs discovered by the Sloan Digital Sky Survey (SDSS), and it has also been observed by the Hubble Space Telescope Advanced Camera for Surveys (HST/ACS; \citealt{Sti05}; \citealt{Kim09}), by the HST Wide Field Camera 3 (HST/WFC3; PI Simcoe, unpublished), and by the Very Large Telescope Multi-Unit Spectroscopic Explorer (VLT/MUSE; ESO archive). Its field is part of the Multiwavelength Yale-Chile survey (MUSYC; \citealt{Gaw06}), which provides imaging in \textit{UBVRIzJHK} down to $B = 26$ and $K = 23$ AB (\citealt{Quad07}), and has also been entirely observed by \textit{Spitzer} IRAC down to 22.5 AB mag at 4.5 $\mu m$ (\citealt{Annunziatella18}).
Near-IR spectroscopy (ISAAAC/VLT) showed that SDSS J1030+0524 is powered by a BH with mass of $1.4\times10^9\;M_{\odot}$ (derived from the Mg\,\textsc{ii} emission line; \citealt{Kur07}; \citealt{DeR11}). Deep and wide optical and near-IR imaging observations of the region ($\sim25'\times25'$) around the QSO with LBT/LBC and CHFT/WIRCam (corresponding to a region of $8\times8$ Mpc at $z=6.31$) also showed that this field features the best evidence to date of an overdense region around a $z\sim6$ QSO (\citealt{Mors14}; \citealt{Bal17}).
The main goals of our deep \textit{Chandra} observation of J1030 were the following: i) to obtain one of the highest quality spectrum ever achieved in the X-rays for a QSO at $z\sim6$ (see \citealt{Nan18}), and ii) to perform a deep X-ray survey in a candidate highly biased region of the Early Universe that has excellent multi-band coverage.
These data were used to study the X-ray variability of the $z=6.31$ QSO SDSS J1030+0524 (\citealt{Nan18}), as well as the diffuse emission detected southward the QSO and associated to a galaxy overdensity at $z = 1.7$ (\citealt{Nan18}; \citealt{Gilli19}), and to characterize the obscured AGN in the field (Peca et al. in prep.). In particular, the $z=1.7$ overdensity is composed by seven galaxy members (six of whom are star-forming) around a central Compton-thick FRII radio source, whose eastern radio lobe is laying at the center of the diffuse X-ray emission and is likely promoting the star formation of the nearby overdensity galaxy members (\citealt{Gilli19}). All these considerations makes J1030 a legacy field for the study of large scale structures around distant accreting SMBHs.
Based on the multi-wavelength coverage of the field, we here present multi-wavelength identifications and basic multi-wavelength photometry for the detected X-ray sources, and their optical/IR counterparts.

The paper is organized as follows. In \S \ref{data_reduction} we describe the \textit{Chandra} data, and the data reduction procedure. In \S \ref{detection} we report the X-ray source detection procedure with a detailed description of the analysis of source completeness and reliability. In \S \ref{catalog}, we present the main X-ray source catalog, and provide the X-ray sources characterization and multi-wavelength identifications. In \S \ref{logNlogS}, we present the cumulative number counts for the main source catalog, and in \S \ref{conclusions} we provide a summary of the main results. Throughout this paper we assume $H_0 = 70$ km s$^{-1}$ Mpc$^{-1}$, $\Omega_{\Lambda} = 0.7$, and $\Omega_M = 0.3$ (\citealt{Ben13}), and errors are reported at 68\% confidence level if not specified otherwise. Upper limits are reported at the 3$\sigma$ confidence level.

%-------------------------------------------------------------------

\section{Observations and Data reduction}\label{data_reduction}

The SDSS J1030+0524 field was observed by \textit{Chandra} with ten different pointings between January and May 2017 for a total exposure of $\sim$479 ks. Observations were taken in the \textit{vfaint} mode for the event telemetry format, using the Advanced CCD Imaging Spectrometer (ACIS) instrument with a roll-angle of $\sim$64\textdegree$\:$ for the first five observations and a roll-angle of $\sim$259\textdegree, for the others. The ten observations (hereafter ObsIDs) cover a total area of roughly 335 arcmin$^2$ and the exposure times of the individual observations range from 26.7 to 126.4 ks. A summary of the observations is provided in Table \ref{tab:general_info}.
\begin{table*}
  \centering
  \captionsetup{justification=centering, labelsep = newline}
      \caption[]{SDSS J1030+0524 observation log}
      \begin{adjustbox}{center, max width=\textwidth}
         \begin{tabular}{c c c c c c}
            \hline
            \hline \rule[0.7mm]{0mm}{3.5mm}
            ObsID & Date & $\theta^{a}$ & $t_{exp}^{b}$ & \multicolumn{2}{c}{Aim Point}\\
            \hhline{~~~~--}
            & & [\textdegree] & [ks] & $\alpha$ (J2000) & $\delta$ (J2000) \\
%            (1) & (2) & (3) & (4) & (5) & (6) & (7) & (8) & (9)\\
            \hline \rule[0.7mm]{0mm}{3.5mm}
            18185 & 2017 Jan 17 & 64.2 & 46.3  & 10 30 28.35 & +05 25 40.2 \\
            \rule[0.7mm]{0mm}{3.5mm}
            19987 & 2017 Jan 18 & 64.2 & 126.4  & 10 30 28.35 & +05 25 35.3\\   
            \rule[0.7mm]{0mm}{3.5mm}
 	   18186 & 2017 Jan 25 & 64.2 & 34.6  & 10 30 28.35 & +05 25 35.3\\   
	   \rule[0.7mm]{0mm}{3.5mm}
  	   19994 & 2017 Jan 27 & 64.2 & 32.7 & 10 30 28.35 & +05 25 37.6\\  
	    \rule[0.7mm]{0mm}{3.5mm}
  	   19995 & 2017 Jan 27 & 64.2 & 26.7 & 10 30 28.35 & +05 25 34.2\\ 
	   \rule[0.7mm]{0mm}{3.5mm}
  	   18187 & 2017 Mar 22 & 259.2 & 40.4 & 10 30 26.67 & +05 24 07.1\\ 
	   \rule[0.7mm]{0mm}{3.5mm}
 	   20045 & 2017 Mar 24 & 259.2 & 61.3 & 10 30 26.66 &  +05 24 07.5\\
	   \rule[0.7mm]{0mm}{3.5mm}
  	   20046 & 2017 Mar 26 & 259.2 & 36.6 & 10 30 26.56 &  +05 24 13.3\\
	   \rule[0.7mm]{0mm}{3.5mm}
	   19926 & 2017 May 25 & 262.2 & 49.4 &10 30 26.68 & +05 24 14.2\\
	   \rule[0.7mm]{0mm}{3.5mm}
 	   20081 & 2017 May 27 & 262.2 & 24.9 & 10 30 26.66 & +05 24 15.2\\[2pt]
            \hline 
         \end{tabular}
        \end{adjustbox}
        \begin{tablenotes}
        		\footnotesize 
		\item \begin{enumerate}[(a)]
			\item Roll-angle in degrees of the ACIS-I instrument.
			\item Exposure time after background flare removal.
		\end{enumerate}
	 \end{tablenotes}
	 \label{tab:general_info}
\end{table*}  

The data were reprocessed using the \textit{Chandra} software CIAO v. 4.8. Data analysis was carried out using only the events with ASCA grades 0, 2, 3, 4, and 6. We then produced X-ray images in the soft, hard, and full bands for each ObsID.

\begin{figure}
 \centering
 \includegraphics[height=8.4cm, width=9cm]{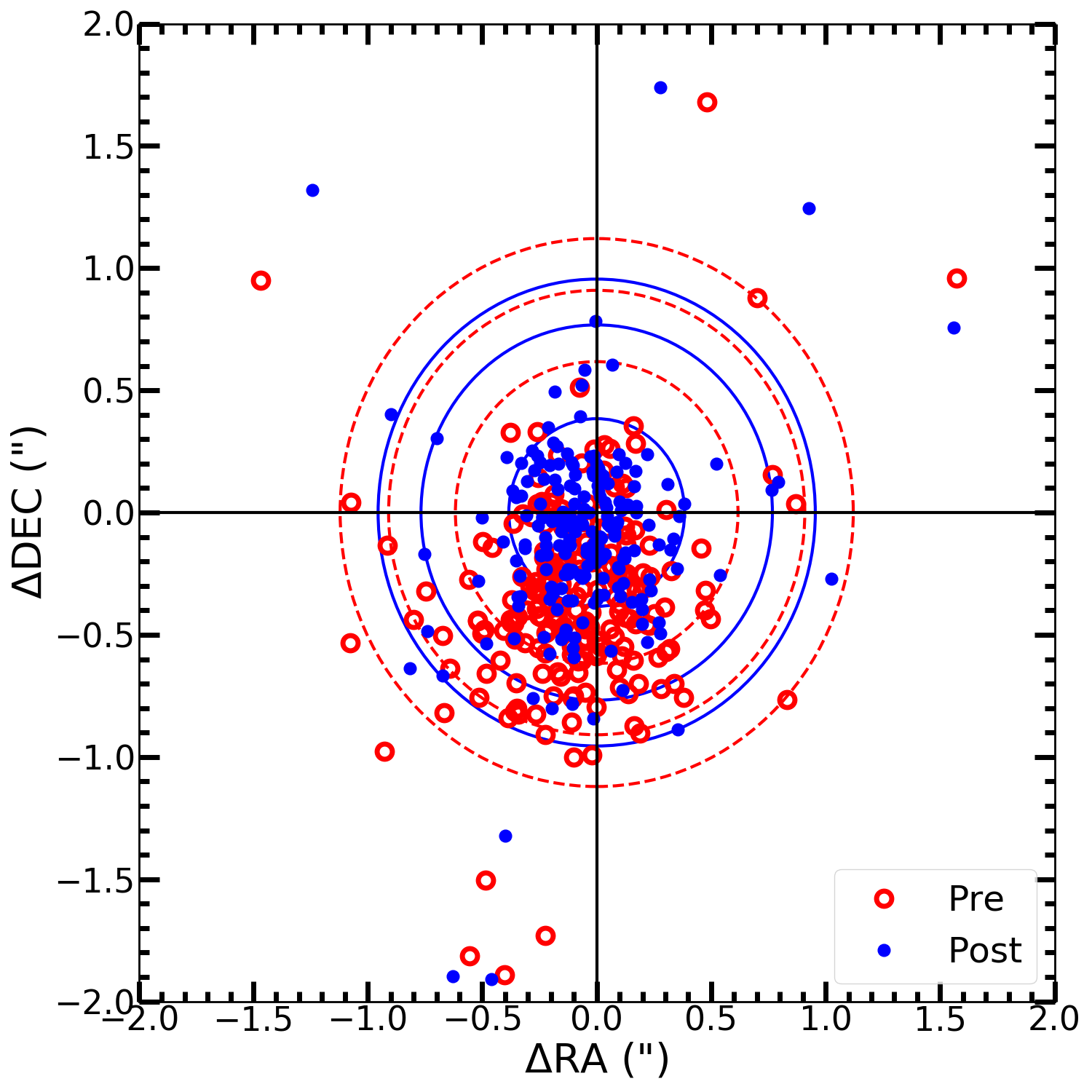}
 \caption{X-ray to $J$-band separation ($\Delta$RA, $\Delta$DEC) in arcsec for X-ray sources detected in each single observation (with a \textit{wavdetect} false-positive probability detection threshold set to 10$^{-6}$ and off-axis <$6'$; see \S \ref{data_reduction} for the details) before (red open circles) and after (blue solid circles) the astrometric correction. The circles encompass 68\%, 90\%, and 95\% of the sources before (red dashed line) and after (blue solid line) the astrometric correction.}
 \label{fig:separation_correction}
\end{figure}

After the data reduction, we corrected the astrometry (applying shift and rotation corrections) of the individual ObsIDs using as reference the WIRCam catalog, which contains $J$-band selected sources down to $J_{AB}=23.75$ (\citealt{Bal17}). First, we created exposure maps and point spread function (PSF) maps for all ObsIDs using the CIAO tools \textit{fluximage} and \textit{mkpsfmap}, respectively. The exposure and PSF maps were computed for the 90\% of the encircled energy fraction (EEF) and at an energy of 2.3, 1.4, and 3.8 keV for the full, soft, and hard band, respectively.
Then, we ran the \textit{Chandra} source detection task \textit{wavdetect} (\citealt{Free02}) on the $0.5-7$ keV images to detect sources to be matched with the $J$-band detected objects.
We set the false-positive probability detection threshold to a conservative value of 10$^{-6}$ and used a ``$\sqrt{2}$ sequence'' of wavelet scales up to 8 pixels (i.e., 1.41, 2, 2.83, 4, 5.66, and 8 pixels) in order to detect only the brightest sources with a well-defined X-ray centroid.
For the match we considered only 43 X-ray sources with a positional error\footnote{Computed as: $\sqrt{\sigma_{RA}^2+\sigma_{Dec}^2}$, where $\sigma_{RA}$ and $\sigma_{Dec}$ are the errors on Right Ascension and Declination, respectively, from \textit{wavdetect}.} below $\sim$0.4" and off-axis <$6'$. We used the CIAO tool \textit{wcs\_match} and \textit{wcs\_update} to match these 43 sources and correct the astrometry, and create new aspect solution files. We considered a matching radius of 2" and applied both translation and rotation corrections. The new aspect solutions were then applied to the event files and the detection algorithm was run again (using the same \textit{wavdetect} parameters and criteria previously adopted). The applied astrometric correction reduces the mean angular distance between the X-ray sources and their $J$-band counterparts from $\theta=0.253$" to $\theta=0.064$". 
As shown in Fig. \ref{fig:separation_correction}, we found that, after applying the astrometric corrections, the distance ($d$) between the X-ray sources used for the astrometric correction and the optical counterparts is $d<0.38'',0.77'',0.95''$ for 68\%, 90\% and 95\% of the X-ray sources, respectively (to be compared with $d<0.62'',0.91'',1.12''$ before the correction). Despite the astrometric corrections, a mean offset of $\Delta RA=-0.07\pm0.3$ and $\Delta DEC=-0.1\pm0.4$ is still present. 
We performed several tests, changing both the off-axis angles and the full band net counts cuts, to verify whether the offset is due to a particular source (or a group of them), but the offset persists and is consistent with the values reported.
However, this offset unlikely affects the matching analysis described in \S \ref{identification}, as the X-ray sources positional errors are generally larger.

\begin{figure}
 \centering
 \includegraphics[height=10cm, width=10cm, keepaspectratio]{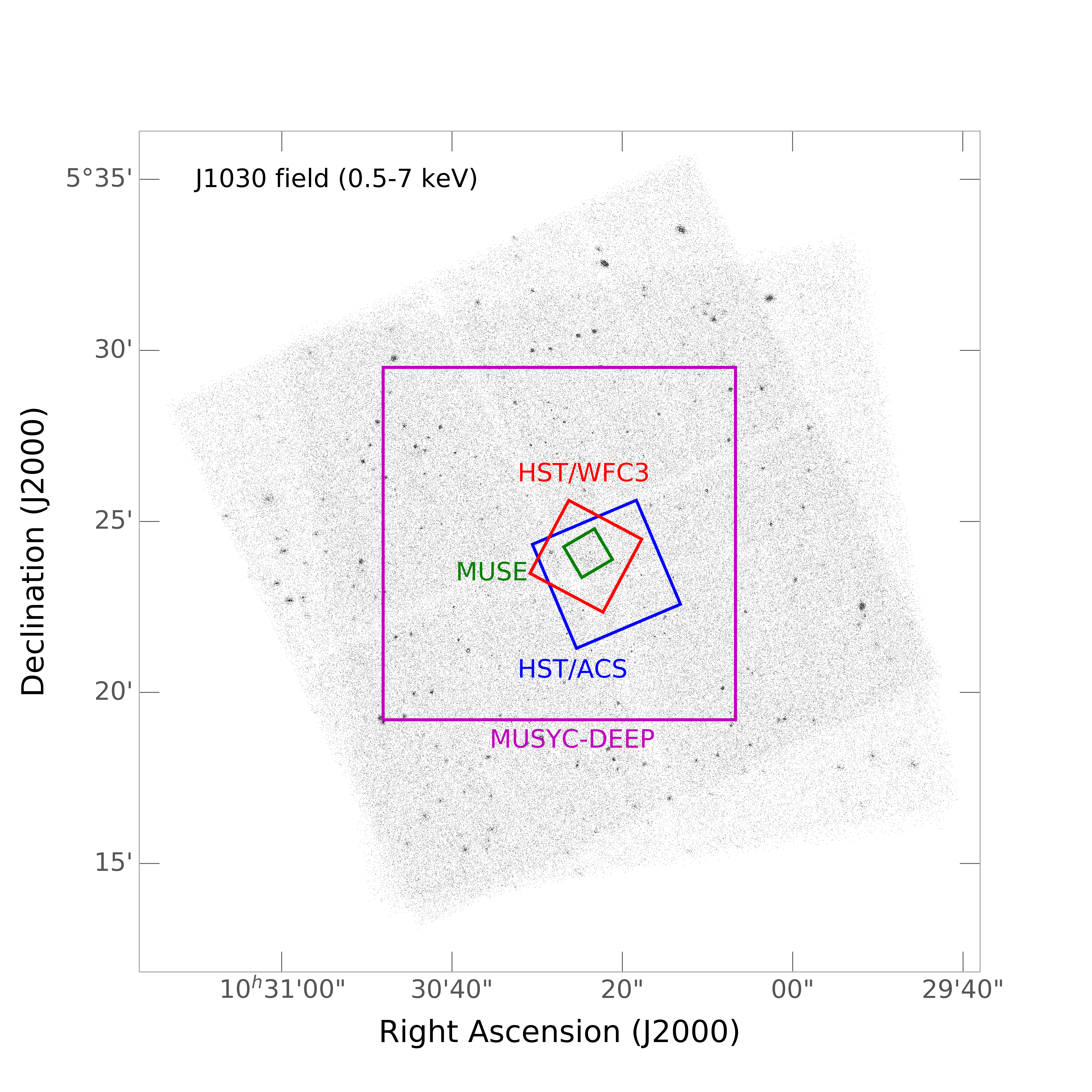}
 \caption{Full-band ($0.5-7$ keV) \textit{Chandra} ACIS-I image of the SDSS J1030+0524 field in logarithmic gray scale. The multi-color regions show some of the central multi-wavelength coverage of the field: the MUSYC-DEEP in purple, HST/WFC3 in red, HST/ACS in blue, and VLT/MUSE in green.}
 \label{fig:j1030_field}
\end{figure}

\begin{figure*}
 \centering
 \includegraphics[height=18cm, width=18cm, keepaspectratio]{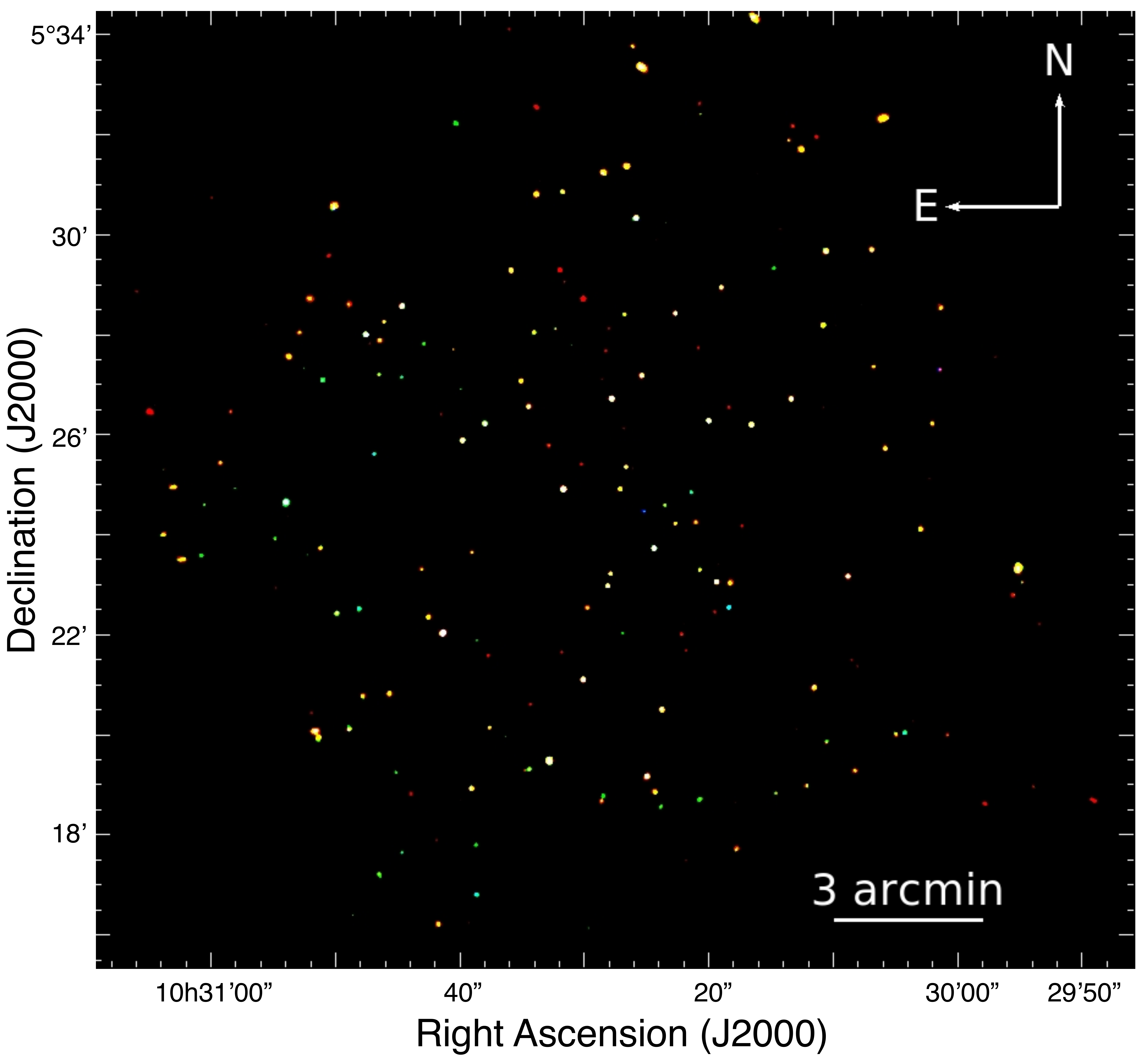}
 \caption{Smoothed ``false-color'' image of the SDSS J1030+0524 field. Colors correspond to 0.5-2.0 keV (red), 2-4.5 keV (green), and 4.5-7 keV (blue).}
 \label{fig:colored_field}
\end{figure*}

\begin{figure}
 \centering
 \includegraphics[height=9.4cm, width=9.4cm, keepaspectratio]{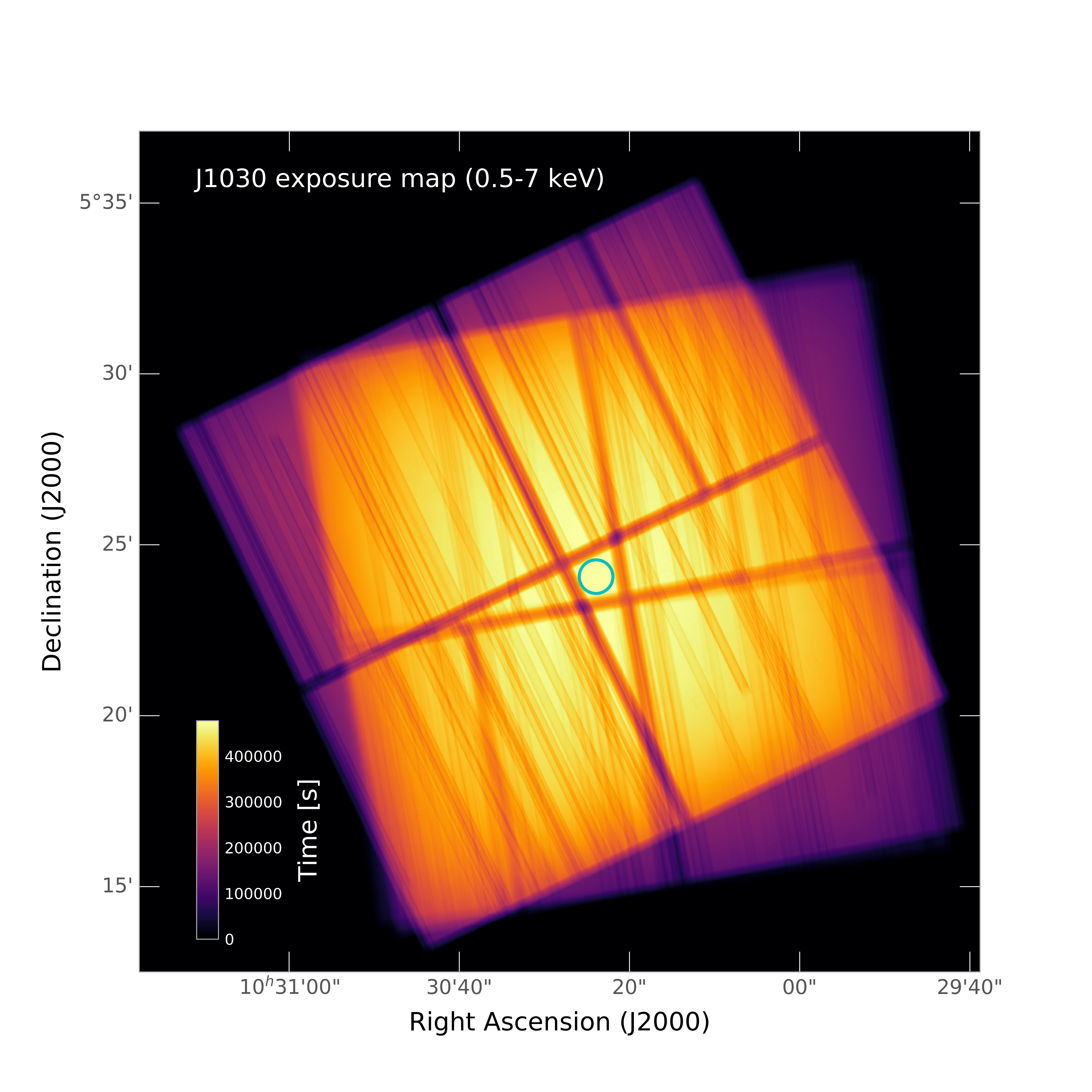}
 \caption{Full-band ($0.5-7$ keV) \textit{Chandra} effective-exposure map of the SDSS J1030+0524 field. The linear scale color bar is shown in the bottom left; the displayed effective exposure times are in units of s. The cyan circle ($r=30''$) marks the position of the $z=6.31$ QSO SDSS J1030+0524.}
 \label{fig:expmap}
\end{figure}

Finally, we stacked the corrected event files using the \textit{reproject\_obs} task and created X-ray images from the merged event file using the standard ASCA grade set in the full, soft, and hard bands. In Fig. \ref{fig:j1030_field} we display the final \textit{Chandra} full-band image with the coverage of the innermost multi-wavelength fields mentioned in \S \ref{intro}. A false-color X-ray image of the field is shown in Fig. \ref{fig:colored_field}. The individual PSF maps were combined using the task \textit{dmimgcalc} to return the exposure-weighted average PSF value at each pixel location in the combined mosaic, while the individual effective-exposure maps were summed together to obtain the total effective-exposure map of the field in the full, soft, and hard bands. The full-band effective-exposure map is shown in Fig. \ref{fig:expmap}.

\section{X-ray source detection}\label{detection}

The X-ray source detection procedure follows a two-stage approach that has also been adopted in past deep X-ray surveys such as the CDF-S (i.e., \citealt{Xue11}; \citealt{Luo17}), and the CDF-N (\citealt{Xue16}): a preliminary list of source candidates was initially generated by \textit{wavdetect} source detection, and then filtered after photometry performed with ACIS Extract (AE; \citealt{Broos12}) to produce our final source catalog.

\subsection{Generation of the preliminary catalog}\label{preliminary_catalog}

To generate the preliminary candidate source list, we ran \textit{wavdetect} on the merged images in the full, soft, and hard bands, using a $\sqrt{2}$ sequence of wavelet scales up to 16 pixels (i.e., 1.414, 2, 2.828, 4, 5.656, 8, 11.314, and 16 pixels) and a false-positive probability threshold of $10^{-4}$. We also provided \textit{wavdetect} with the average PSF maps (\S2.1) for each energy band. This produced 498, 383, 370 candidate sources in the full, soft, and hard bands, respectively. Among them, 289, 221, and 218 sources are also detected in the full, soft, and hard bands, respectively, when running \textit{wavdetect} with a more conservative threshold of $10^{-5}$ that is used in many deep \textit{Chandra} surveys.
The loose \textit{wavdetect} source-detection threshold of $10^{-4}$ is expected to introduce a large number of spurious detections that must be filtered out, but also allows us to push the detection to the faintest possible limits.

We then improved the source positions through the AE ``CHECK\_POSITIONS'' procedure and used AE to extract photometric properties of the candidate sources. The details of the AE photometric extraction are described in the AE User's Guide, and a summary is also provided in \citet{Xue11}. We used AE to perform source and background extractions for each source in each ObsID and then we merged the results. In our case, a polygonal extraction region that approximates the $\sim$90\% encircled energy fraction contour of the local PSF, at $E=1.4$ keV in the full and soft bands, and at $E=2.3$ in the hard band, was utilized to extract source counts. We adopted the AE ``BETTER\_BACKGROUNDS'' algorithm for background extraction (see \S 7.6.1 of the AE User's Guide), in order to obtain a single background region plus a background scaling that simultaneously models all background components, including the background that arises from the PSF wings of neighboring sources. A minimum number of 100 counts in the merged background spectrum is required to ensure photometric accuracy, which was achieved through the AE ``ADJUST\_BACKSCAL'' stage. The extraction results from individual observations were then merged to produce photometry for each source through the AE ``MERGE\_OBSERVATIONS'' stage.
To filter the preliminary catalog, the most important output parameter from AE is the binomial no-source probability ($P_B$), which is the probability of observing at least the same number of source counts under the assumption that there is no real source at that location and that the observed number of counts is purely due to a background fluctuation:
\begin{equation}
P_B(X\ge S)= \sum_{X=S}^N\frac{N!}{X!(N-X)!}p^X(1-p)^{N-X},
\label{eq:binomial}
\end{equation}
where $S$ is the total number of counts in the source extraction region (before background subtraction); $N = S + B_{ext}$, where $B_{ext}$ is the total number of counts in the background extraction region; and $p = 1/(1 + BACKSCAL)$ with $BACKSCAL = A_{ext}/A_{src}$ is the ratio between the background and source extraction regions. We computed $P_B$ for each source in all the three (full, soft, hard) bands.
Although $P_B$ is a classic confidence level, usually it is not a good indicator of the fraction of spurious sources (e.g., a cut at $P_B = 0.01$ does not correspond to a 1\% spurious rate), mainly because the extractions were performed on a biased sample of candidate sources that already survived a filtering process by \textit{wavdetect}. Furthermore, given its definition, the value of $P_B$ is dependent on the choice of source and background extraction regions. Therefore, we cannot reject spurious sources simply based on the absolute value of $P_B$ itself. A $P_B$ threshold derived from simulations needs to be adopted to maximize the completeness and reliability of our sample.

\subsection{Generation of the simulated data}

To clean the catalog from spurious sources as much as possible and to assess the completeness and reliability of our final sample we produced three simulations that closely mirror our observations. A similar procedure has been already used in previous X-ray surveys (e.g., \citealt{Cappe07}; \citeyear{Cappe09}; \citealt{Pucc09}; \citealt{Xue11}; \citeyear{Xue16}; \citealt{Luo17}).

First, we considered a mock catalog of X-ray sources (AGN and normal galaxies) that covers an area of one square degree and reaches fluxes that are well below the detection limit of our $\sim$479 ks exposure. 
In this mock catalog, we assigned to each simulated AGN a soft-band flux randomly drawn from the soft-band log(N)-log(S) relation expected in the AGN population synthesis model by \citet{Gilli07}. Simulated galaxy fluxes were drawn randomly from the soft-band galaxy log(N)-log(S) relation of the ``peak-M'' model of \citet{Ran05}. 
The AGN and galaxy integrated flux is consistent within the uncertainties with the cosmic X-ray background flux (CXB; see e.g., \citealt{Cappelluti17}).
AGN and galaxies have been simulated down to $2\times10^{-18}$ erg/cm$^2$/s in the 0.5-2 keV band, to include the contribution of undetectable sources that produce the spatially non-uniform background component. The soft-band fluxes of the simulated AGN and galaxies were converted into full-band fluxes assuming power-law spectra with $\Gamma = 1.4$\footnote{This value is typically used to translate count rates into fluxes for the AGN population, since it describes fairly well the observed slope of the CXB and is therefore representative of a sample that includes both unobscured and obscured AGN (\citealt{Hick06}).} and $\Gamma = 2.0$, respectively. 
The number of simulated sources has been rescaled for the J1030 field area ($\sim$335 arcmin$^2$), and source coordinates were randomly assigned within that area. We used the MARX software (v. 5.3.3; \citealt{Dav12}) to convert source fluxes to a Poisson stream of dithered photons and to simulate their detection by ACIS-I.

Second, we produced ten simulated ACIS-I observations of the mock catalog, each configured to have the same aim point, roll-angle, exposure time, and aspect solution file of each of the ten J1030 pointings (see Table \ref{tab:general_info}). These simulated source event files contain the actual number of photons produced on the detector by our simulated sources. We then produced the corresponding background event files from the J1030 event files. For each real event file, we masked all the events associated to our preliminary source candidates and then filled the masked regions with events that obey the local probability distribution of background events (using the \textit{blanksky} and \textit{blanksky\_sample} tools). The simulated source event files produced using MARX were then merged with the background event files to produce ten simulated ACIS-I pointings that closely mirror the ten real ones.

As a final step, we followed the same approach adopted above: 
\begin{itemize}
\item We stacked the ten simulated event files using the \textit{reproject\_obs} task and created X-ray images in all the three X-ray bands from the merged event file (\S \ref{data_reduction}).
\item We ran \textit{wavdetect} on each simulated combined image at a false-positive probability threshold of $10^{-4}$ to produce a catalog of simulated source candidates, and used AE to perform photometry (including the $P_B$ values) on them (\S \ref{preliminary_catalog}).
\end{itemize}
The procedure above was repeated three times, allowing us to generate a total of three complete simulations that mirror the J1030 field and to obtain a simulated preliminary source catalog from each simulation.

\subsection{Completeness and reliability}\label{completeness_reliability}

Simulations are the best tools to set a probability threshold ($P_B$) for source filtering since we have full control of the input and output sources.
After creating the three simulations and the corresponding candidate source catalogs (as described in \S3.2), we matched the detected output simulated sources with the input sources in the mock catalog, using a likelihood-ratio matching technique (see e.g., \citealt{Cili03}; \citealt{Brusa07}).
The goal of this method is to distinguish, among the detected sources in the simulation, between input simulated sources and spurious ones. Once the detected sources are classified, we adopted other filters based on the X-ray properties of each source (like the $P_B$; see the detailed description reported below) that can be applied also to our \textit{Chandra} observation.

Briefly, our likelihood-ratio (LR) technique takes into account the positional accuracy of the output sources, and also the expected flux distribution of the input ones. It assigns likelihood and reliability parameters to all possible counterparts, and mitigates the effect of false matches.
For an input source with a flux $f$ at an angular separation $r$ from a given output source, the LR is the ratio between the probability of the input being the true counterpart of the output and the corresponding probability of the input of being an unrelated (i.e., background) object (e.g., \citealt{SutSan92}; \citealt{Brusa05}):
\begin{equation}
 LR = \frac{F(r)q(f)}{n(f)}
  \label{eq:ciliegi2003}
\end{equation}
where $F(r)$ is the probability distribution function of the angular separation, $q(f)$ is the expected flux distribution of the input sources, and $n(f)$ is the surface density of background objects with flux $f$. We refer to Appendix A for a complete explanation of Equation \ref{eq:ciliegi2003}. In our case, for each output source we searched for input sources inside a circular area of $r_{LR}=5"$ (following \citealt{Luo10}) centered on the output position, to allow matching of the X-ray sources at large off-axis angles.
\begin{comment}
We assumed that $F(r)$ follows a Gaussian distribution:
\begin{equation}
 F(r) = \frac{1}{2 \pi \sigma^2}exp\Bigg(\frac{-r^2}{2\sigma^2}\Bigg)
\end{equation}
where $\sigma = PSF_{Radius}/\sqrt{C}$ (taken from \citealt{Pucc09}), and C are the net, background-subtracted, counts computed by AE, and $PSF_{Radius}$ is evaluated with the estimate at the 90\% encircled energy radius (at E = 1.5 keV) given by Equation 1 from \citet{Hick06}.
\end{comment}

A threshold for LR (LR$_{th}$) is needed to discriminate between spurious and real associations: an output source is considered to have an input counterpart if its LR value exceeds LR$_{th}$. The choice of LR$_{th}$ depends on two factors: first, LR$_{th}$ should be small enough to avoid missing many real identifications, so that the sample completeness is high; second, LR$_{th}$ should be large enough to keep the number of spurious identifications low, in order to increase the reliability of the identifications.
The reliability of a single input source $j$, which represents the probability of being the correct identification, is defined as:
\begin{equation}
R_j = \frac{LR_j}{\sum_i LR_i+(1-Q)}
\end{equation}
where the sum is over all the possible input counterparts for an output source $i$, and $Q=\int_{f_{lim}}^{+\infty} q(f)df$ is the probability that the input counterpart $j$ is brighter than the flux limit ($f_{lim}$) of the catalogue. We then defined the reliability parameter (R) for the total sample as the ratio between the sum of the reliabilities of all sources identified as possible counterparts and the total number of sources with LR > LR$_{th}$ ($N_{LR > LR_{th}}$):
\begin{equation}
R = \frac{\big(\sum_j R_j\big)_{LR>LR_{th}}}{N_{LR > LR_{th}}},
\end{equation}
We also measure the completeness parameter (C) of the total sample defined as the ratio between the sum of the reliability of all sources identified as possible counterparts and the total number of output sources ($N_X$):
\begin{equation}
C = \frac{\big(\sum_j R_j\big)_{LR>LR_{th}}}{N_{X}}.
\end{equation}
As proposed in \citet{Brusa07} and \citet{Civ12}, $LR_{th}$ was computed as the likelihood-ratio which maximizes the quantity (R + C)/2 (we found $LR_{th}=3.61,4.05,3.31$ for the full, soft, and hard band, respectively). We hence flagged those sources with LR > LR$_{th}$ as good matches and those with LR < LR$_{th}$ as spurious matches.

\begin{figure}
 \centering
 \includegraphics[height=9cm, width=9cm, keepaspectratio]{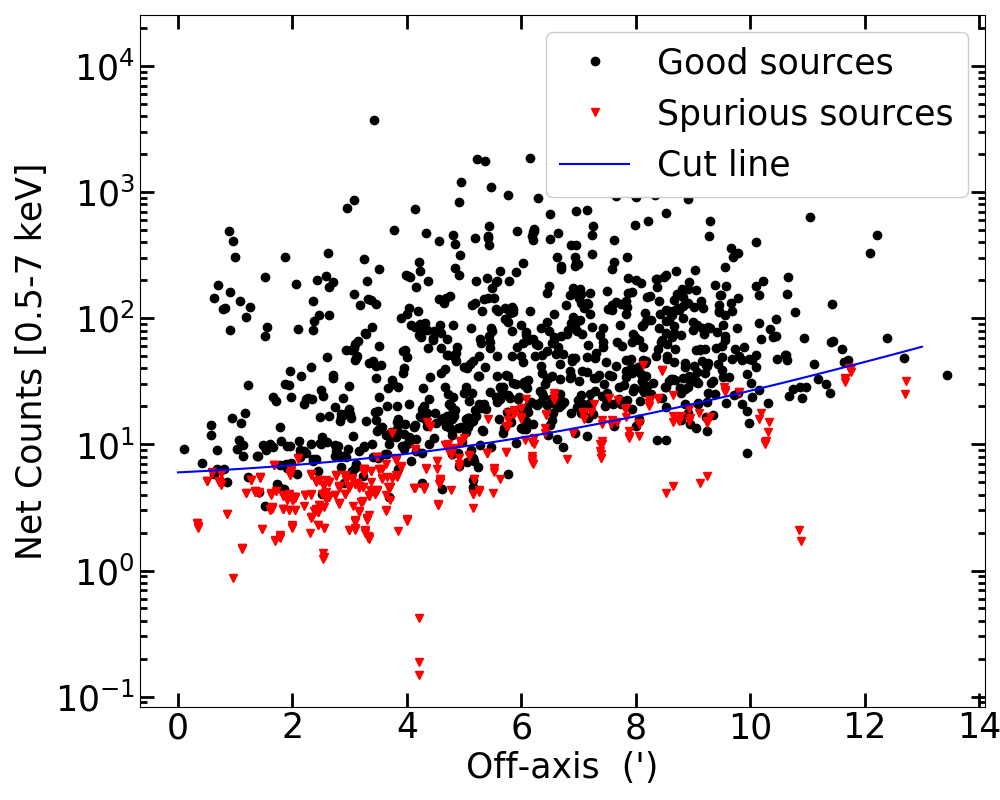} \includegraphics[height=9cm, width=9cm, keepaspectratio]{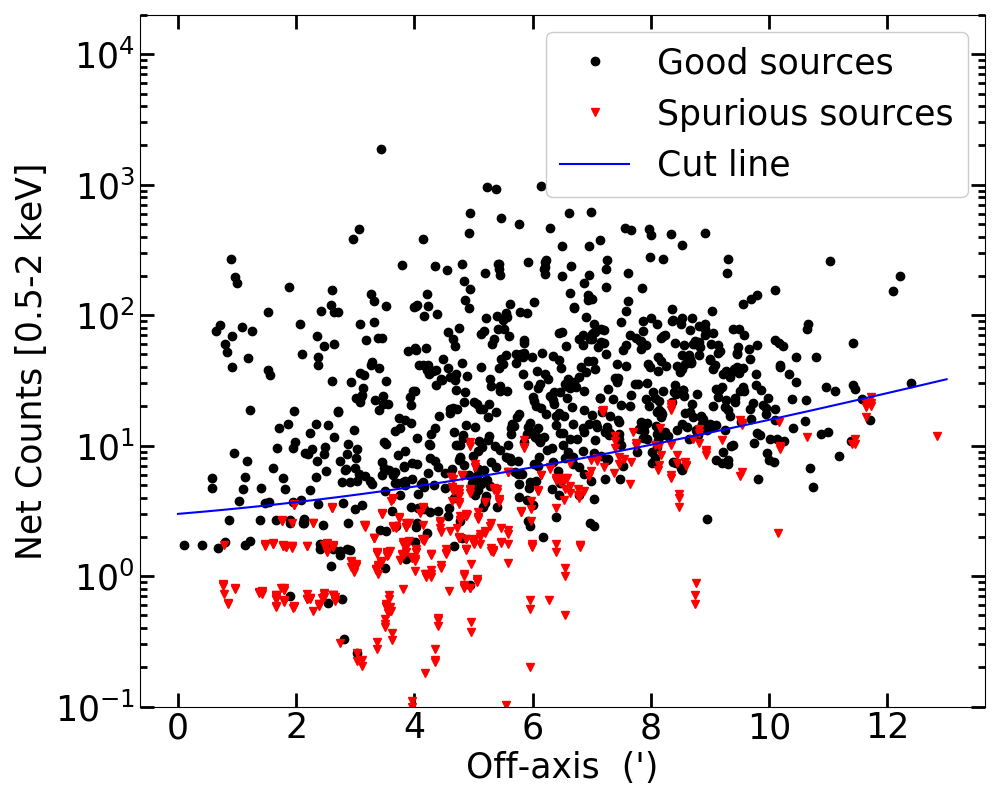} \includegraphics[height=9cm, width=9cm, keepaspectratio]{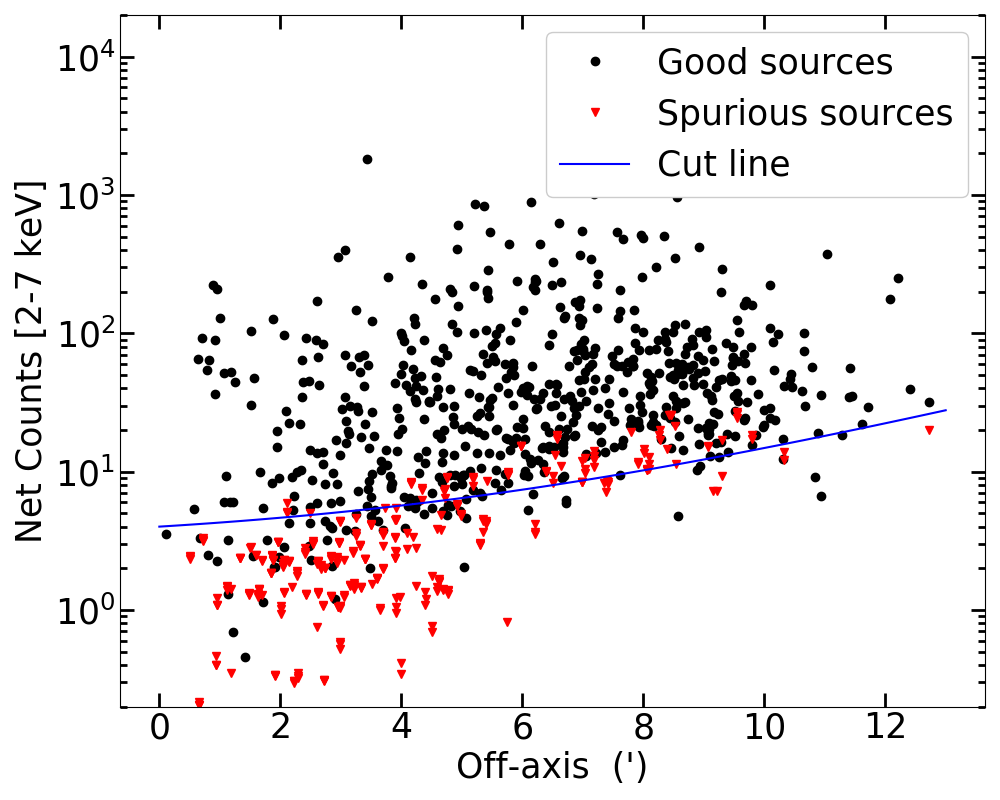}
\caption{Net counts vs off-axis angles for the sources detected in the three simulations in the full (top), soft (middle), and hard (bottom) bands. The black dots are the ``good'' input-output matches according to the likelihood-ratio method, while the red triangles are those considered spurious. The blue line represents the source-count limit as a function of the off-axis angle that was adopted to compute the completeness and reliability.}
\label{fig:net_cts_vs_offaxis}
\end{figure}

We used those matches flagged as ``good'' to assess the completeness and reliability of the simulated catalog. 
Looking at the distributions of the net counts in the three bands (full, soft, and hard) vs off-axis angle for both good and spurious matches, we derived three empirical linear cut relations (one for each band) that define an effective source-count limit as a function of the off-axis angles. These cut lines (blue lines in Fig. \ref{fig:net_cts_vs_offaxis}) maximize the number of rejected spurious sources while keeping the number of rejected ``good'' sources around $10\%$.
Then, by considering only those sources above the cut lines, we defined the completeness as the ratio between the number of ``good'' sources detected with a binomial probability above a certain value and the total number of ``good'' sources. The reliability is defined as 1 minus the ratio between the number of spurious sources above a certain probability value and the total number of sources.

\begin{figure*}
 \centering
 \includegraphics[height=6cm, width=6cm, keepaspectratio]{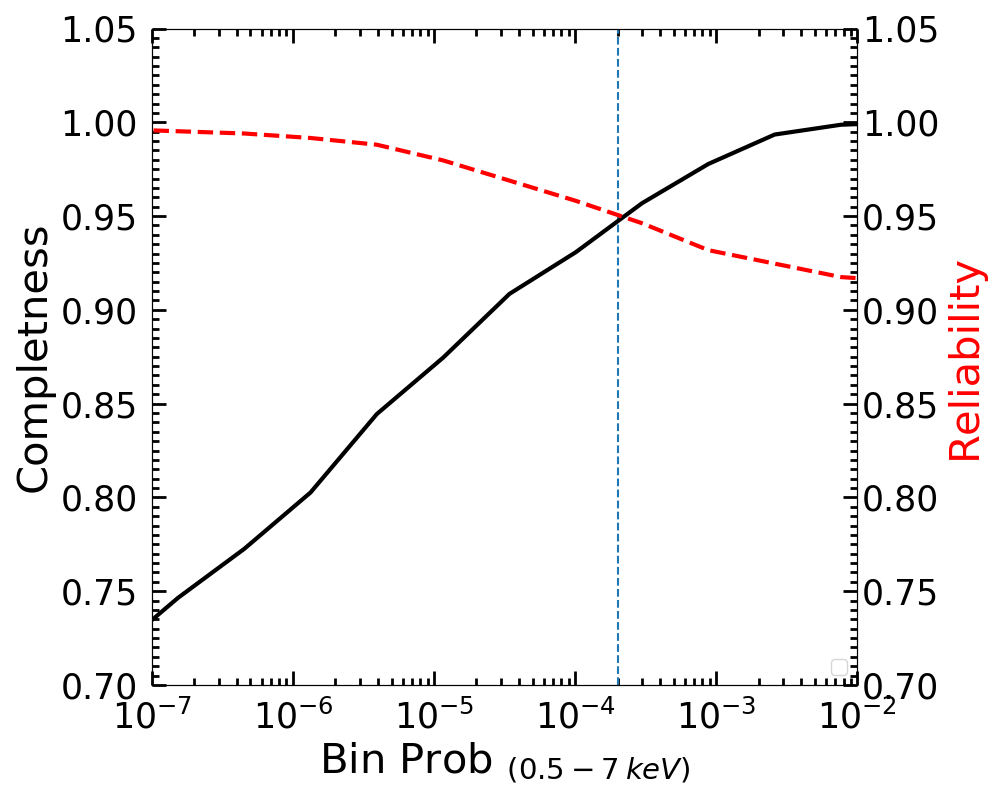} \includegraphics[height=6cm, width=6cm, keepaspectratio]{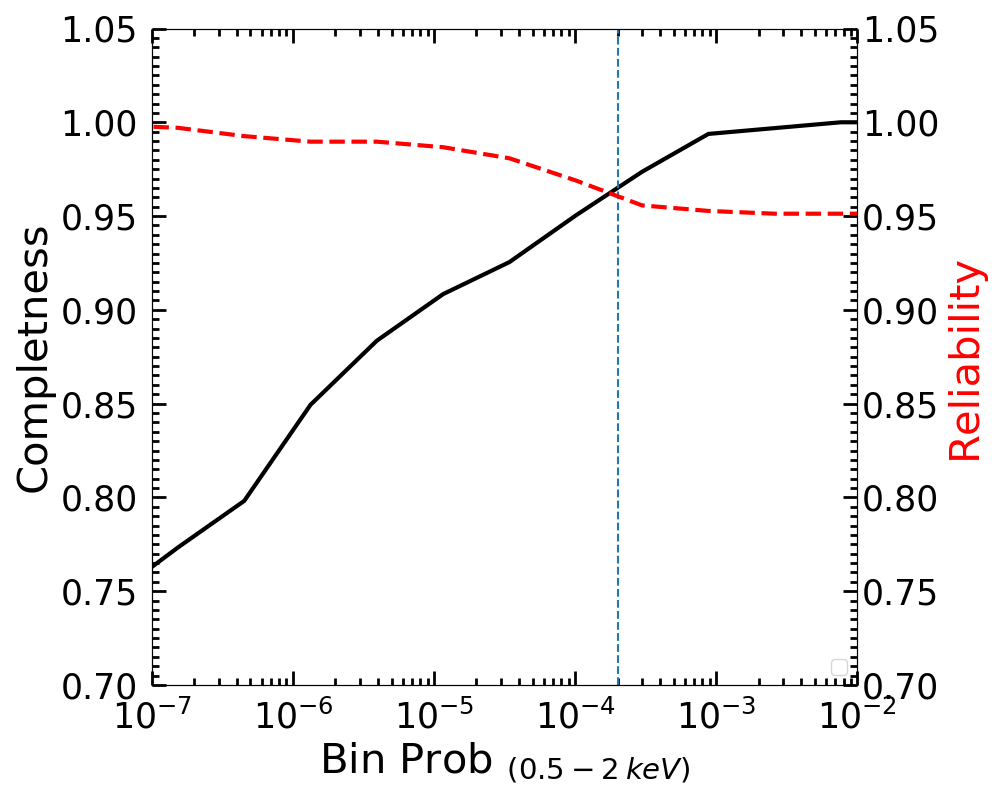} \includegraphics[height=6cm, width=6cm, keepaspectratio]{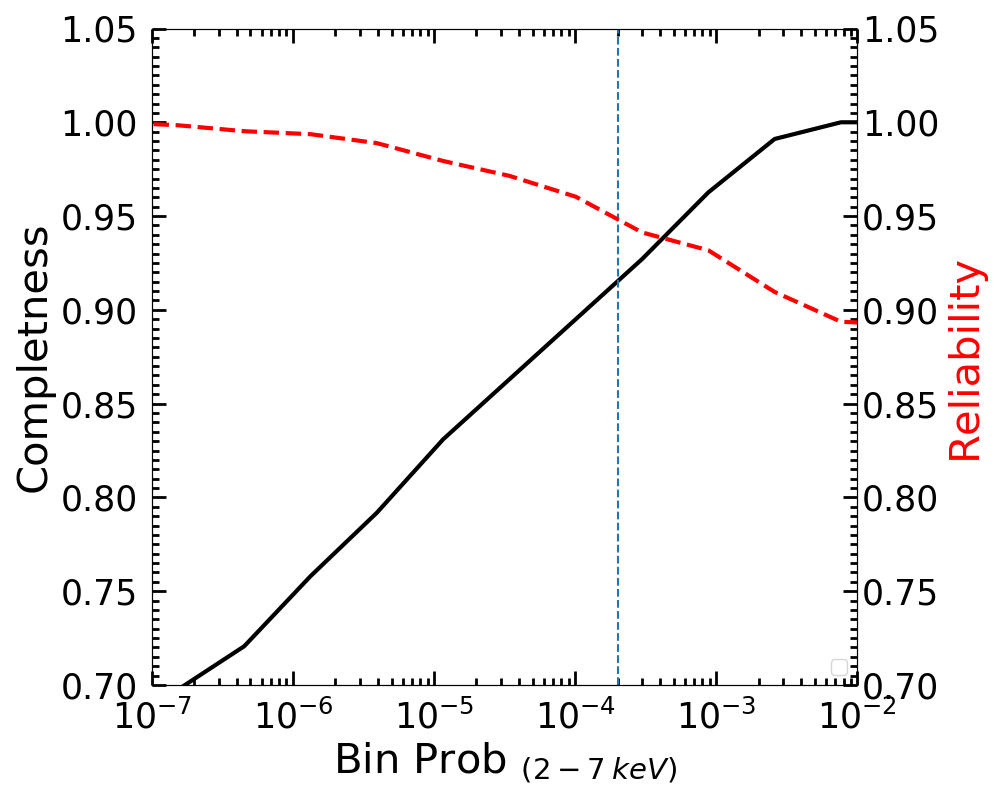}
 \caption{Completeness (C, black curve) and reliability (R, red dashed curve) as a function of the binomial no-source probability for the three combined simulations in the full (left), soft (middle), and hard (right) bands, respectively. The dashed vertical line indicates the chosen source-detection threshold of $P_{B} = 2\times10^{-4}$.}
 \label{fig:completeness_reliability}
\end{figure*}

In Fig. \ref{fig:completeness_reliability} we show the completeness (black solid line) and reliability (red dashed line) as a function of the $P_B$ for the simulations in the full, soft, and hard bands, for sources that survived our count-limit relation cuts. We adopted a probability threshold value of $P_B=2\times10^{-4}$ to keep the reliability values $>95\%$ in all the three bands. In fact, adopting $P_B=2\times10^{-4}$, the completeness levels for the entire J1030 field are 95\% (full band), 97\% (soft band), and 91\% (hard band), while the reliability levels are 95\% (full band), 96\% (soft band), and 95\% (hard band). Finally, we applied our source-count limit cut and the $P_{B}=2\times10^{-4}$ threshold derived from the simulations as filters for our J1030 preliminary real source catalog: 244, 193, and 208 sources survived in the full, soft, and hard bands, respectively. Compared to past X-ray surveys, it is the first time that more sources are detected in the hard band rather than in the soft one, and we ascribed it to the rapid degradation of the soft-band \textit{Chandra} effective area that occurred in the last few years.

We then matched the three band catalogs using a matching radius $r=5"$ (for $5''<r<10''$ no additional sources are matched), and visually checked all matches. The final catalog contains 256 unique sources, detected in at least one band.

We also computed the completeness and reliability (following the same approach described above) for source catalogs generated with a \textit{wavdetect} false-positive probability threshold of $10^{-5}$. Based on simulations, we verified that catalogs generated with \textit{wavdetect} threshold $10^{-5}$ and ($P_B=10^{-3}$) have the same reliability and completeness and similar source numbers of those obtained by using $10^{-4}$ and $P_B=2\times10^{-4}$.
After a visual check of those sources that are not in common between the two catalogs, we found that the catalog obtained with a \textit{wavdetect} threshold of $10^{-4}$ and $P_B=2\times10^{-4}$ contains more associations with optical/IR counterparts than the other one, and we hence adopt it as our final X-ray source catalog.

\section{Source catalog}\label{catalog}

As explained in \S3.3, our final catalog consists of 256 sources detected in one or more X-ray bands (full, soft, and hard); in Table \ref{tab:band_statistic} we report the total number of sources for each band combination.
\begin{table}
\centering
  \captionsetup{justification=centering, labelsep = newline}
      \caption[]{Number of \textit{Chandra} sources detected in up to three bands}
      \begin{adjustbox}{center, max width=\textwidth}
         \begin{tabular}{c c}
            \hline
            \hline \rule[0.7mm]{0mm}{3.5mm}
            Band & Number of \\
            \hhline{~~}
           $[$keV$]$ & sources \\
%            (1) & (2) & (3) & (4) & (5) & (6) & (7) & (8) & (9)\\
            \hline \rule[0.7mm]{0mm}{3.5mm}
            F+S+H & 156 \\
            \rule[0.7mm]{0mm}{3.5mm}
            F+S & 31\\   
            \rule[0.7mm]{0mm}{3.5mm}
 	   F+H & 46\\
	   \rule[0.7mm]{0mm}{3.5mm}
 	   F & 11\\
	   \rule[0.7mm]{0mm}{3.5mm}
 	   S & 6\\
	   \rule[0.7mm]{0mm}{3.5mm}
 	   H & 6\\
	   \rule[0.7mm]{0mm}{3.5mm}
 	   \textbf{Total} & \textbf{256}\\
	   [2pt]
            \hline 
         \end{tabular}
        \end{adjustbox}
        \begin{tablenotes}
        		\footnotesize
		\item The bands reported are the full (F), soft (S), and hard (H).
	\end{tablenotes}
	\label{tab:band_statistic}
\end{table} 
Eleven sources are detected only in the full-band, 6 only in the soft-band, and 6 only in the hard-band.
For each source we derived the net counts, hardness ratio (HR), and band fluxes and relative errors or upper limits (for those sources that are not detected in a given band), as described in \S \ref{xray_properties}. In Table \ref{tab:cts_statistic} we report the basic statistics of the source counts in the three bands.
\begin{table}
  \centering
  \captionsetup{justification=centering, labelsep = newline}
      \caption[]{Statistics of \textit{Chandra} detected sources}
      \begin{adjustbox}{center, max width=\textwidth}
         \begin{tabular}{c c c c c c}
            \hline
            \hline \rule[0.7mm]{0mm}{3.5mm}
            Band  & Number of & \multicolumn{4}{c}{Net counts per source}\\
            \hhline{~~----}
           $[$keV$]$ & sources & Max & Min & Mean & Median\\
%            (1) & (2) & (3) & (4) & (5) & (6) & (7) & (8) & (9)\\
            \hline \rule[0.7mm]{0mm}{3.5mm}
            Full & 244 & 1849.3 & 6.7 & 96.5 & 50.9\\
            \rule[0.7mm]{0mm}{3.5mm}
            Soft & 193 & 1196.1 & 2.7 & 61.0 & 41.5\\   
            \rule[0.7mm]{0mm}{3.5mm}
 	   Hard & 208 & 647.6 & 3.4 & 53.3 & 40.7\\[2pt]
            \hline 
         \end{tabular}
        \end{adjustbox}
	\label{tab:cts_statistic}
\end{table}  

\subsection{X-ray properties}\label{xray_properties}

\begin{figure}
 \centering
 \includegraphics[height=15.4cm, width=9.cm, keepaspectratio]{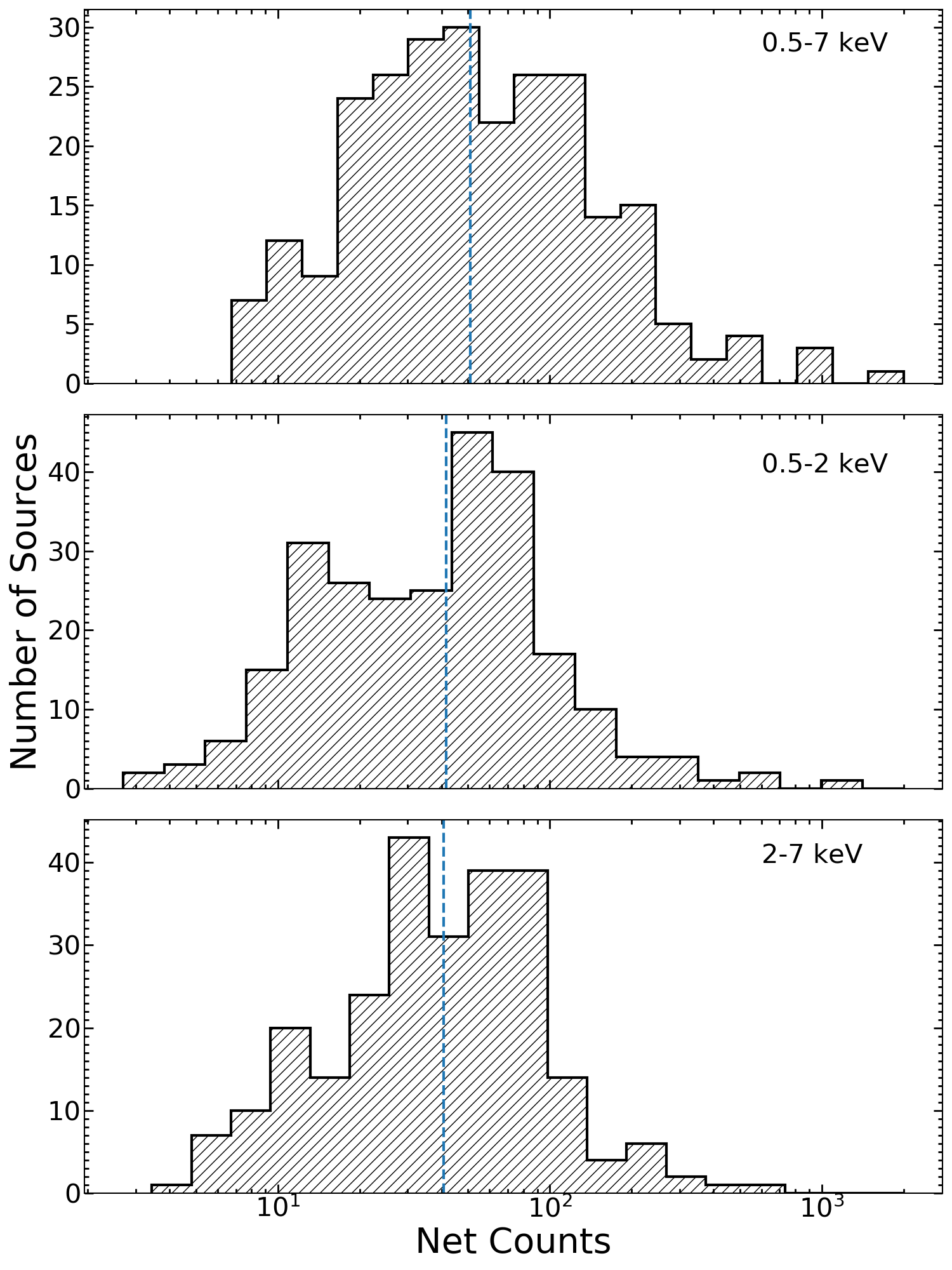}
 \caption{Net counts distributions for the \textit{Chandra} sources detected in the full (top), soft (middle), and hard (bottom) bands. The cyan dotted vertical lines mark the medians of the distributions: 50.9, 41.5, and 40.7 net counts for the full, soft, and hard bands, respectively. Sources with upper limits on the counts are not included in these plots.}
 \label{fig:cts_distribution}
\end{figure}

In each of the three X-ray bands, the net source counts were derived from the AE ``MERGE\_OBSERVATIONS'' procedure using a polygonal extraction region that approximates the $\sim90\%$ of the encircled energy fraction at $E=1.4$ keV in the full and soft bands, and at $E=3.8$ in the hard band, as explained in \S3, while the associated 1$\sigma$ errors are computed by AE following \citet{Geh86}.
For those sources that are below the detection threshold ($P_B > 2\times 10^{-4}$) in one or two bands, we computed the 3$\sigma$ upper limits using the \textit{srcflux} tool of CIAO, that extracts source counts from a circular region, centered at the source position, that contains 90\% of the PSF at 2.3, 1.4, and 3.8 keV in the full, soft, and hard band, respectively. Their counts from the background are extracted from an annular region around the source location, that has an inner radius equal to the size of the source radius and an outer radius five times larger.
The distributions of the source counts in the three bands are displayed in Fig. \ref{fig:cts_distribution}.

We used the net count rates in the different bands to compute the hardness ratio (HR) for each source. The hardness ratio was computed as:
\begin{equation}
HR = \frac{H-S}{H+S},
\label{eq:hr}
\end{equation}
where H and S are the net rates (the ratio between the net counts and the effective exposure time at the source position) in the hard and soft bands, respectively. Errors are computed at the 1$\sigma$ level following the method described in \citet{Lyo91}. Upper and lower limits were computed using the 3$\sigma$ net counts upper limits. For the 11 sources with only full-band detection, we could not compute the HR.

\begin{figure}
 \centering
 \includegraphics[height=15.4cm, width=9cm, keepaspectratio]{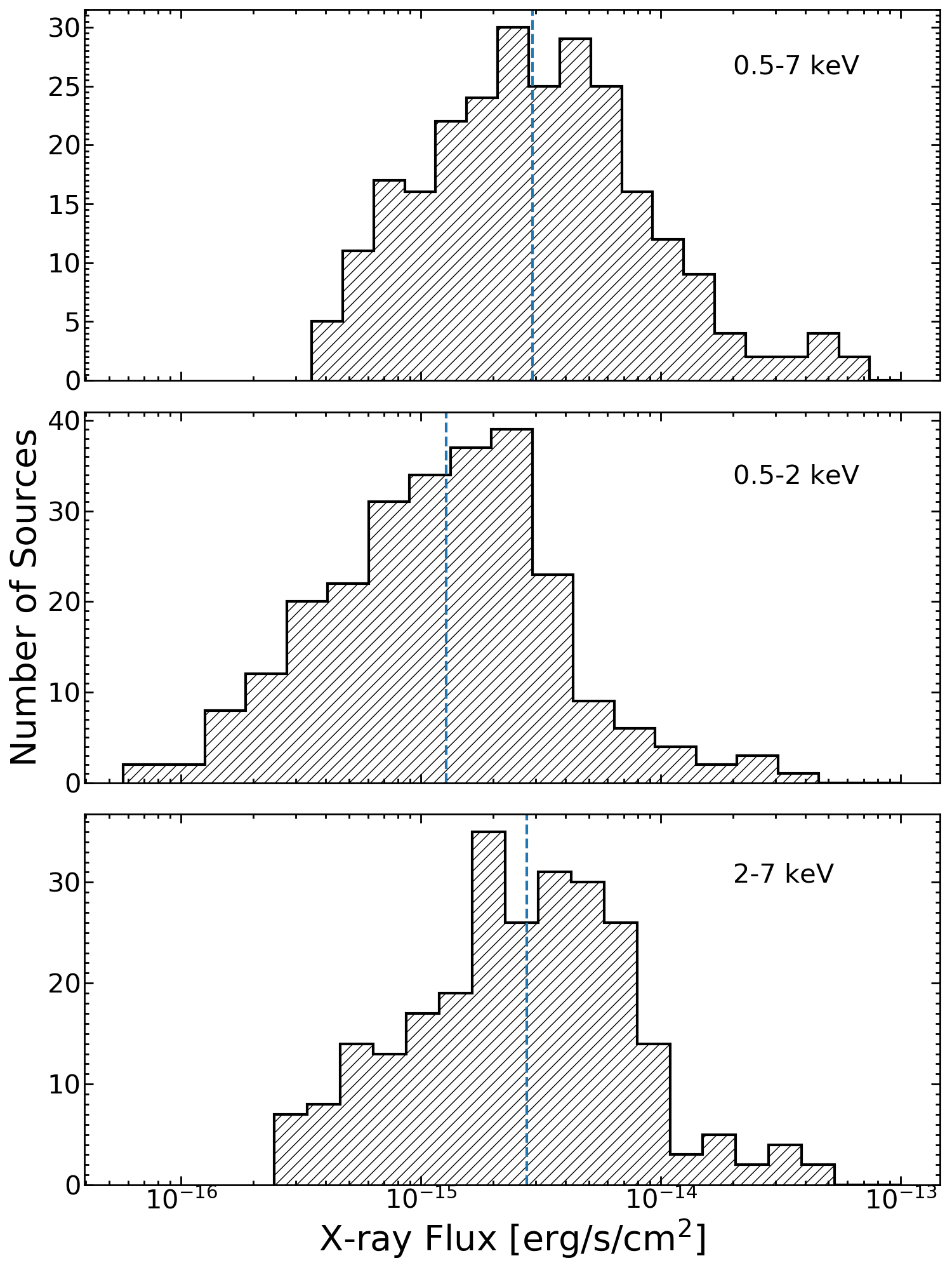}
 \caption{Aperture corrected X-ray flux distributions for the sources detected in the full (top), soft (middle), and hard (bottom) bands. The cyan dotted vertical lines mark the medians of the distributions: $2.9, \,1.3,\, 2.8 \times 10^{-15}$ erg s$^{-1}$ cm$^{-2}$ for the full, soft, and hard bands, respectively. Sources with upper limits on the counts are not included in these plots.}
 \label{fig:flux_distribution}
 \end{figure}

While a detailed spectral analysis of the X-ray sources is beyond the scope of the current study, we converted the aperture corrected count rates (or their upper limits) to the corresponding fluxes (or flux upper limits) in a given band assuming that their spectra are power-laws modified by only Galactic absorption ($N_H=2.6\times10^{20}$ cm$^{-2}$) with effective power-law photon indices derived from the hardness ratios. At fixed $N_H$ and redshift, the HR is a function of the power-law index (e.g., see Fig. 10 in \citealt{Mar16}). The HR-$\Gamma$ relation was derived simulating 100 X-ray spectra with $N_H=0$, $z=0$, and different $\Gamma=-2-+2$ (steps of 0.1), and deriving the corresponding HRs.
For the sources not detected in both the soft and hard bands, the hardness ratios cannot be constrained, so we assumed a spectral power-law with $\Gamma=1.4$ (i.e., the mean value derived for the slope of the CXB; \citealt{Hick06}) modified by Galactic absorption. In this case, the adopted count rate-to-flux conversion factors are $CF= CR/flux =6.3,9.8,4.4\times10^{10}$ counts erg$^{-1}$ cm$^2$ for the full, soft, and hard bands, respectively.
The distributions of the source fluxes in the three bands are displayed in Fig. \ref{fig:flux_distribution}.
      
\subsection{Multi-wavelength Source Identifications}\label{identification}

We searched for optical and IR counterparts of the X-ray sources in our LBT/LBC, CHFT/WIRCam, and \textit{Spitzer} IRAC ($4.5\,\mu m$ band) catalogs (see \citealt{Mors14}, \citealt{Bal17}, and \citealt{Annunziatella18}, respectively), using a likelihood-ratio matching technique similar to that described in \S \ref{completeness_reliability}. Again, a threshold value for the likelihood-ratio that maximizes the (R+C)/2 value was chosen ($LR_{th}=1.06,1.71,2.41,1.11$ for the $r$, $z$, $J$, and IRAC bands, respectively). For the X-ray sources with multiple counterpart candidates that satisfy our likelihood threshold, we selected the candidate with the highest reliability level. In particular, we found 17, 7, and 9 X-ray sources that have multiple counterpart candidates that satisfy the likelihood threshold in the $r$, $z$, and $J$ bands, respectively, while there are no multiple counterpart candidates in the IRAC $4.5\,\mu m$ band.
%The unassigned counterpart candidates have reliabilities $<48\%$, $<49\%$, and $<18\%$ in the $r$, $z$, and $J$ bands, respectively.

For the optical and IR identifications, we used the following four catalogs:
\begin{itemize}
\item  The J1030+0524 LBC $z$ and $r$ bands catalogs, that contain 29150 and 86150 sources with limiting AB magnitudes of 25.2 and 27.5, respectively (50\% completeness limit; \citealt{Mors14}). In \citet{Mors14} the $z$-band data were used as master images on which object detection (5$\sigma$) was made, then the measurements were performed on the $r$-band images only to obtain spatially coherent photometric colors. Subsequently, we performed a source detection on the deeper $r$ image to produce an independent $r$-band catalog with a limiting AB magnitude of 27.5.
\item The J1030+0524 Wide-field InfraRed Camera (WIRCam) \textit{J}-band (NIR) catalog that contains 14770 sources down to $J_{AB} = 23.75$ (50\% completeness limit at $5\sigma$; \citealt{Bal17}).
\item The J1030+0524 \textit{Spitzer} IRAC at MIR $4.5\,\mu m$ band (MIR) catalog that contains 16317 sources down to $m_{4.5\mu m} = 22.5$ (50\% completeness limit at $5\sigma$; \citealt{Annunziatella18}).
\end{itemize}

\begin{figure}
 \centering
 \includegraphics[height=21cm, width=10cm, keepaspectratio]{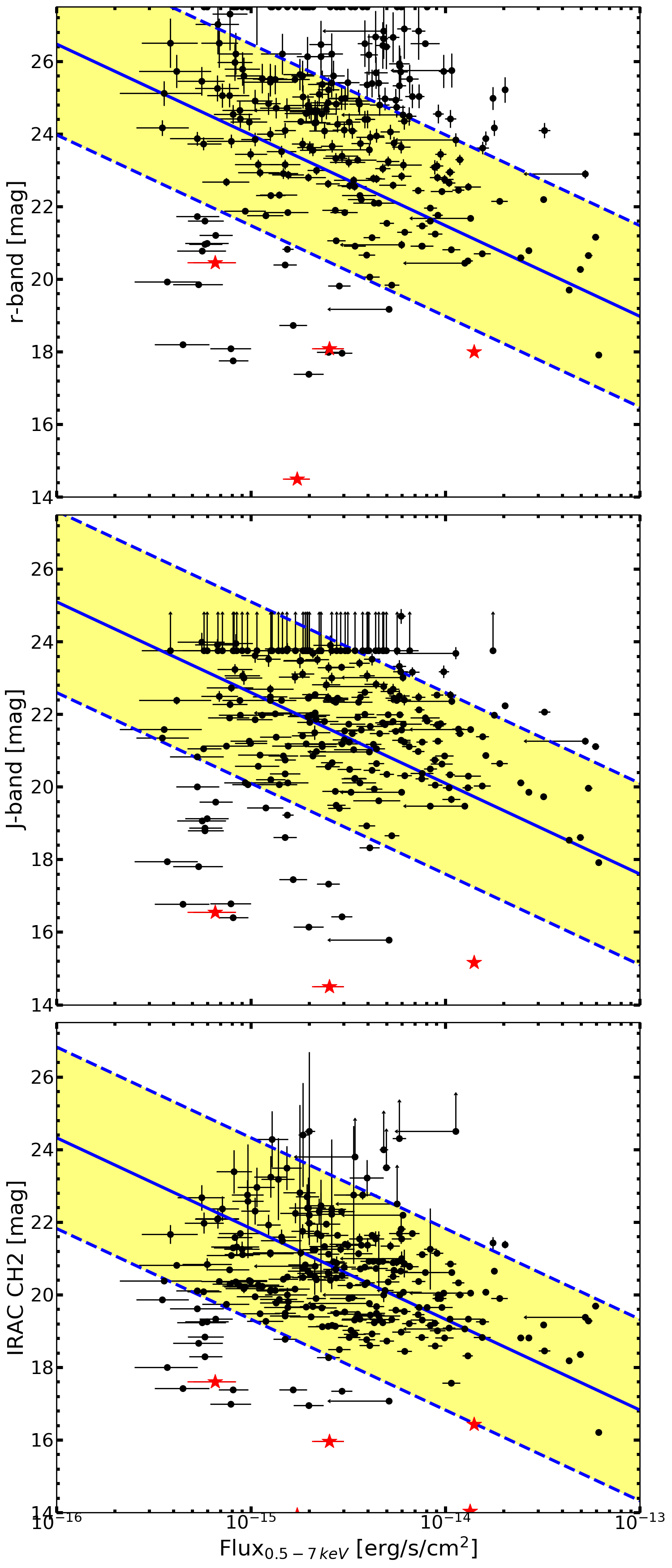}
 \caption{X-ray full-band flux vs. $r$-band (top panel), $J$-band (middle panel), and $4.5\mu m$-band (bottom panel) AB magnitudes. Black dots represent the main catalog sources, while red stars represent the known stars of our catalog. The blue diagonal dashed lines show constant X-ray to $r$-, $J$-, or $4.5\mu m$-band flux ratios $log(f_X/f_{opt})=-1,+1$, while the blue diagonal solid line shows $log(f_X/f_{opt})=0$. The yellow shaded region highlights the area between the blue diagonal dashed lines, that for the optical bands represents the ``classic locus'' of unobscured AGN.}
 \label{fig:flux_mag}
 \end{figure}

We initially identified unique counterparts for 244 (95.3\%) of the 256 main-catalog sources. We examined the 12 X-ray sources that lack counterparts, and assigned multi-wavelength matches to eight sources (with off-axis angle $>1'$ and $LR$ below but close to $LR_{th}$) for which the X-ray centroid computed by AE is too far away ($>1''$) from the most likely optical counterpart to provide a LR value above our adopted threshold, while the positional error is consistent with the optical counterpart. %Since the reliability of our catalog is 95\%, we expect to have $\sim13$ sources (5\%) that are spurious among the 21 that have no optical counterpart.
After this adjustment, we then obtained primary counterparts in at least one optical/NIR/MIR band for 252 (98.4\%) of the 256 main-catalog sources. 
Among these 252 X-ray sources there are 1, 6, and 6 sources that have $r$, $z$, or $J$ band counterparts, respectively, that are below the limiting magnitude of the corresponding survey. For these sources, we performed aperture photometry to obtain the missing catalog band magnitudes but we did not compute the corresponding likelihood and reliability, as we have no information on the magnitude distribution of  background sources at such faint fluxes. %Postage-stamp images for the X-ray catalog sources are reported in Fig. \ref{fig:optical_counterpart}. The images are color composites of the LBT/LBC $r$, $z$, and CFHT/WIRCam $J$ bands. Furthermore, we show in Fig. \ref{fig:irac_counterpart} 17 X-ray sources that have just IRAC counterparts.
The distributions of the X-ray full-band fluxes versus LBT/LBC $r$-band, CHFT/WIRCam $J$-band, and IRAC CH2 $4.5\mu m$-band magnitudes for the main catalog sources are displayed in Fig. \ref{fig:flux_mag}. The blue diagonal lines show constant X-ray to $r$- or $J$-band flux ratios defined \citep[similarly to][]{Civ12} as:
\begin{equation}
log(f_X/f_{opt}) = log(f_X) + C_{opt} + m_{opt}/2.5 = -1,0,+1
\end{equation}
where $f_X$ is the X-ray full-band flux, $m_{opt}$ is the magnitude at the chosen optical/IR band, and $C_{opt}$ is a constant which depends on the specific filter used in the optical observations. Considering the bandwidths and the effective wavelength of the LBC $r$-band, WIRCam $J$-band, and IRAC CH2 $4.5\mu m$-band filters, we used $C_r=5.41$, $C_J=5.96$, and $C_{4.5\,\mu m}=6.27$. The yellow shaded region between the blue diagonal dashed lines ($-1<log(f_X/f_{opt})<+1$) in Fig. \ref{fig:flux_mag} has been adopted as the reference area where unobscured AGN are expected to lay in the optical bands, while obscured AGN are expected at $log(f_X/f_{opt})>1$ (e.g., \citealt{Brusa10}; \citealt{Civ12}). The higher number of sources above the $log(f_X/f_{opt})=1$ relation observed in the $r$-band (Fig. \ref{fig:flux_mag}, upper panel) compared to the $4.5\,\mu m$-band (Fig. \ref{fig:flux_mag}, bottom panel) is probably related to the lower nuclear extinction in the IR that in the optical bands. Red stars represent the X-ray sources identified as stars based on optical information: XID9, XID63, XID146, XID147, and XID162.

Most of the \textit{Chandra} source counterparts have been spectroscopically observed with the LBT Multi-Object Double CCD Spectrograph (MODS) and the Large Binocular Telescope Near-infrared Spectroscopic Utility with Camera and Integral Field Unit for Extragalactic Research (LUCI) for a total of 52 hrs (16 hrs with LUCI and 36 hrs with MODS) to measure their redshifts. The data reduction and analysis are in progress and the derived properties will be released in the next future (Mignoli et al. in prep.). We will also use the dense multi-band coverage in the J1030 field to derive the photometric redshifts of the X-ray sources (Marchesi et al. in prep.). Photometric redshifts will be used whenever the optical/NIR spectroscopy is missing.

%Finally, we noted that there are X-ray sources with multiple counterpart candidates that satisfy our threshold cut and has a reliability >30\%. As mentioned before, these X-ray sources have been matched with the optical counterparts that have the highest reliability level. However, the secondary counterparts (e.i., those with lower reliability level but still aboe 30\%) are presented in Table...

\subsection{Main catalog description}\label{catalog_entries}

We present the main \textit{Chandra} source catalog in Table \ref{tab:j1030_catalog}. The details of the table columns are given below.
\begin{enumerate}
\item Column 1: source sequence number (XID).
\item Columns 2 and 3: right ascension (R.A.) and declination (DEC) of the X-ray source, respectively. These positions were computed through the AE ``CHECK\_POSITIONS'' procedure (\S \ref{preliminary_catalog}). In the catalog we report the centroid position derived from the full-band. For sources not detected in the full-band we used the centroid positions derived either from the soft or from the hard band.
\item Column 4: positional error on the source centroid. This was computed as $\sigma = PSF_{Radius}/\sqrt{C}$ (\citealt{Pucc09}), where $C$ are the net, background-subtracted, counts computed by AE, and $PSF_{Radius}$ is the 90\% encircled energy radius (at E = 1.4 keV) given by Equation 1 of \citet{Hick06}.
\item Column 5: off-axis angle in arcminutes computed as the angular distance between the position of the X-ray source and the average aim point of the J1030 field (10:30:27.50, +05:24:54.0).
\item Column 6: effective exposure time in ks taken from the full-band exposure map.
\item Column 7-9: hardness ratio computed with Equation \ref{eq:hr} and relative errors. Errors are computed at the 1$\sigma$ level following the method described in \citet{Lyo91}. For display purposes, these three columns are grouped in column 7 of Table \ref{tab:j1030_catalog}.
\item Columns 10-18: net counts and relative errors computed by AE in the full (F), soft (S), and hard (H) bands, respectively. Errors are computed according to Table 1 and 2 of \citet{Geh86} and correspond to the 1$\sigma$ level in Gaussian statistics. For those sources that are not detected in a given band, we provide upper limits at the 3$\sigma$ confidence level (see \S \ref{xray_properties}). For display purposes, each band net counts and errors are grouped in column 8, 9, and 10 for the full, soft, and hard band, respectively, of Table \ref{tab:j1030_catalog}.
\item Columns 19-21: binomial no-source probability $P_B$ computed by AE in the full, soft, and hard bands. Only sources with $P_B < 2\times10^{-4}$ in at least one band are included in the catalog.
\item Columns 22-30: aperture-corrected X-ray fluxes and relative errors in the full, soft, and hard bands, respectively, while for undetected sources we report 3$\sigma$ upper limits. Fluxes and relative errors were computed from the net rates and relative errors assuming that the full-band spectra of the X-ray sources are power-laws modified by only Galactic absorption with effective power-law photon indices derived from their hardness ratios. For the sources not detected in the soft and hard bands, we assumed a spectral power-law with $\Gamma=1.4$ modified by Galactic absorption.
\item Columns 31 and 32: right ascension (R.A.) and declination (DEC), respectively, of the optical/IR counterpart. When available, we provide the centroid position from the $z$-band catalog, otherwise we provide the position in other bands following this order of priority: $J$-band, $r$-band, or $4.5\mu m$-band centroid.
\item Column 33: positional offset between the X-ray source and optical counterpart in arcsecs.
%\item Columns 20-21: R.A. and DEC separation  between the X-ray source and optical/IR counterpart in arcsecs.
\item Columns 34-37: counterpart magnitude AUTO in the $r$, $z$, $J$, and $4.5\mu m$ bands, respectively. The reported limits for the undetected counterparts correspond to the limiting AB magnitudes of the corresponding optical/NIR/MIR catalog.
\item Columns 38-41: counterpart magnitude errors in the $r$, $z$, $J$, and $4.5\mu m$ bands, respectively.
%\item Columns 30-33: counterpart likelihood-ratio value in the $r$, $z$, $J$, and $4.5\mu m$ bands, respectively.
%\item Columns 34-37: reliability value of the counterpart in the $r$, $z$, $J$, and $4.5\mu m$ bands, respectively.
\item Column 42: flag providing info on the likelihood of the counterparts: -1 for X-ray sources with no counterpart in any band, 0 for sources with a sub-threshold counterpart, 1 for sources with unique counterpart above likelihood threshold, 2 for sources with two counterparts above threshold (for which we report the counterpart with the highest LR).
\item Column 43: flag notes for the single XID sources: 0 for sources with no morphological information, 1 for sources that have a star as optical/NIR/MIR counterpart based on the optical information, 2 for sources that appear as X-ray extended sources.
%\item Column 38: flag providing info on the counterparts: 0 for X-ray sources with the counterparts found by the likelihood method, 1 for sources without a counterpart, 2 for counterparts manually assigned even if below threshold, 3 when the counterpart is a known star.
\end{enumerate}

The catalog with all the info reported above is publicly available at: \href{http://www.oabo.inaf.it/~LBTz6/1030/chandra_1030}{http://www.oabo.inaf.it/$\sim$LBTz6/1030/chandra\_1030}.

\begin{table*}
  \begin{center}
  \captionsetup{justification=centering, labelsep = newline}
      \caption[]{\textit{Chandra} source catalog}
      \begin{adjustbox}{center, max width=\textwidth}
         \begin{tabular}{c c c c c c c c c c}
            \hline
            \hline
            XID & R.A. & DEC &  Pos Err $["]$ & Off-axis $[']$ & Exposure $[ks]$ & HR & F & S & H\\
             (1) & (2) & (3) & (4) & (5) & (6) & (7) & (8) & (9) & (10)\\
             \hline \rule[0.7mm]{0mm}{3.5mm}
          1 & 10:30:24.96 & +05:19:09.40 & 0.27 & 5.8 & 391.9 & $-0.12_{-0.08}^{+0.08}$ & $251.7_{-16.3}^{+17.3}$ & $138.6_{-11.9}^{+13.0}$ & $113.0_{-11.1}^{+12.1}$\\
           \rule[0.7mm]{0mm}{3.5mm}
           2 & 10:30:32.82 & +05:19:28.80 & 0.14 & 5.6 & 420.1 & $-0.01_{-0.04}^{+0.04}$ & $827.0_{-29.0}^{+30.0}$ & $416.1_{-20.5}^{+21.5}$ & $420.0_{-20.7}^{+21.8}$\\
           \rule[0.7mm]{0mm}{3.5mm}
          3 & 10:30:23.77 & +05:20:30.23 & 0.24 & 4.5 & 425.4 & $-0.04_{-0.08}^{+0.08}$ & $164.0_{-13.1}^{+14.1}$ & $84.7_{-9.3}^{+10.4}$ & $79.3_{-9.2}^{+10.2}$\\
            \rule[0.7mm]{0mm}{3.5mm}
           4 & 10:30:30.11 & +05:21:05.96 & 0.15  &  3.9 & 428.4 & $-0.15_{-0.07}^{+0.07}$ & $259.4_{-16.2}^{+17.2}$ & $147.6_{-12.2}^{+13.2}$ & $109.9_{-10.6}^{+11.7}$\\
           \rule[0.7mm]{0mm}{3.5mm}
           5 & 10:30:22.19 & +05:22:00.79 &  0.33  &   3.2 &  331.3 & $-0.32_{-0.19}^{+0.19}$ & $36.6_{-6.2}^{+7.3}$ & $23.4_{-4.9}^{+6.0}$ & $12.2_{-3.7}^{+4.8}$\\[2pt]
            \hline
         \end{tabular}
        \end{adjustbox}
         \begin{tablenotes}
        		\footnotesize
		\item A complete version of this table with all the 256 sources and properties listed in \S \ref{catalog_entries} is provided online.
	\end{tablenotes}
	 \label{tab:j1030_catalog}
	 \end{center}
   \end{table*}

%--------------------------------------

\section{Cumulative log(N)-log(S) of the J1030 field}\label{logNlogS}

\begin{figure}
 \centering
 \includegraphics[height=9cm, width=9cm, keepaspectratio]{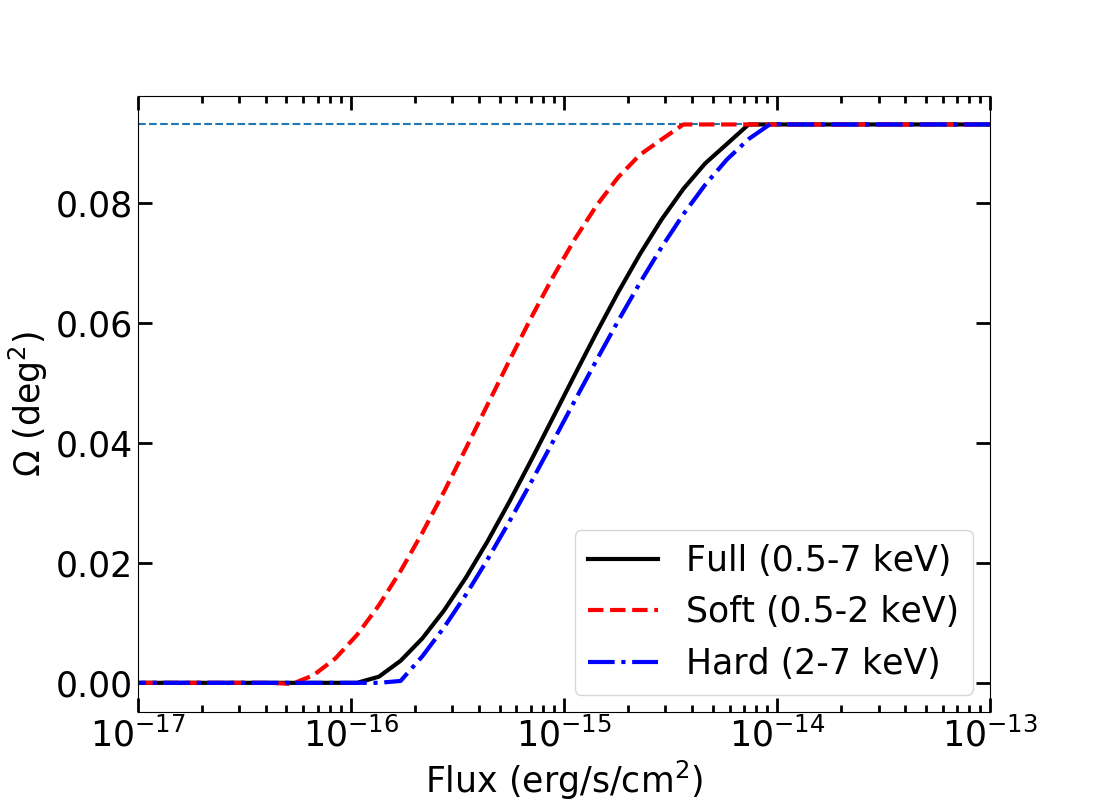}
\caption{Sky coverage (i.e., sky area vs flux limit relation) in the full (black solid line), soft (red dashed line), and hard (blue dash-dotted line) bands derived from our simulations. The horizontal cyan dotted line represents the total geometric area of our J1030 field (335 arcmin$^2$).}
\label{fig:sky_coverage}
\end{figure}

\begin{figure}
 \centering
  \vspace{-0.2cm}\includegraphics[height=8cm, width=9cm, keepaspectratio]{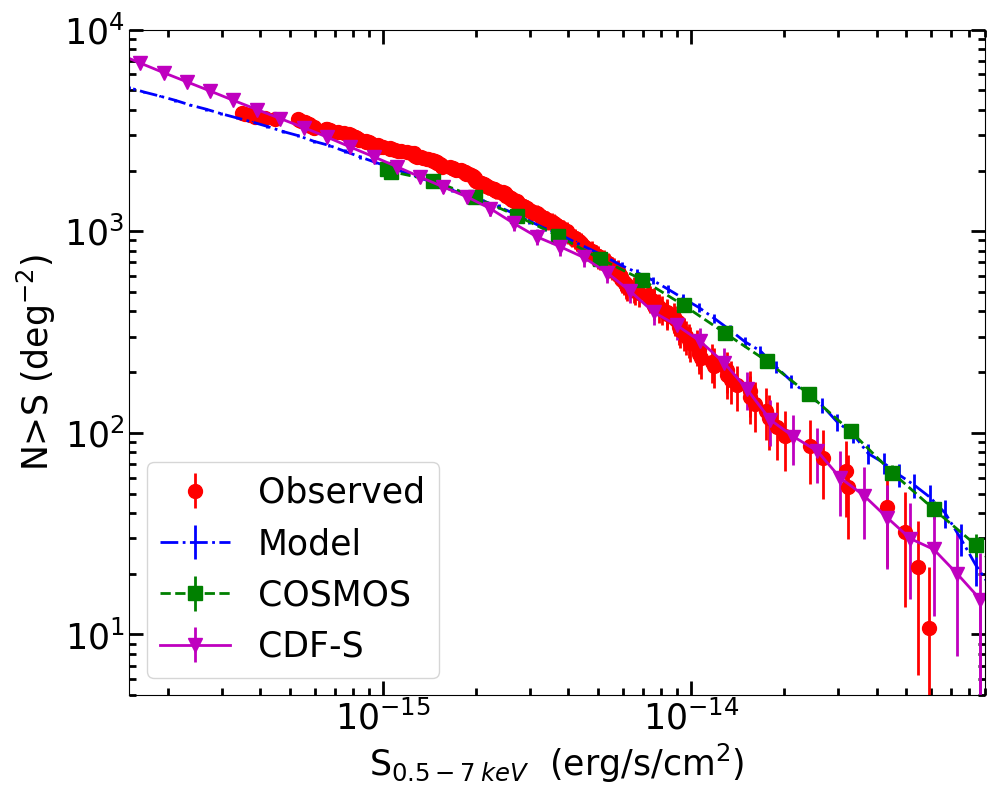} \vspace{-0.05cm} \includegraphics[height=8cm, width=9cm, keepaspectratio]{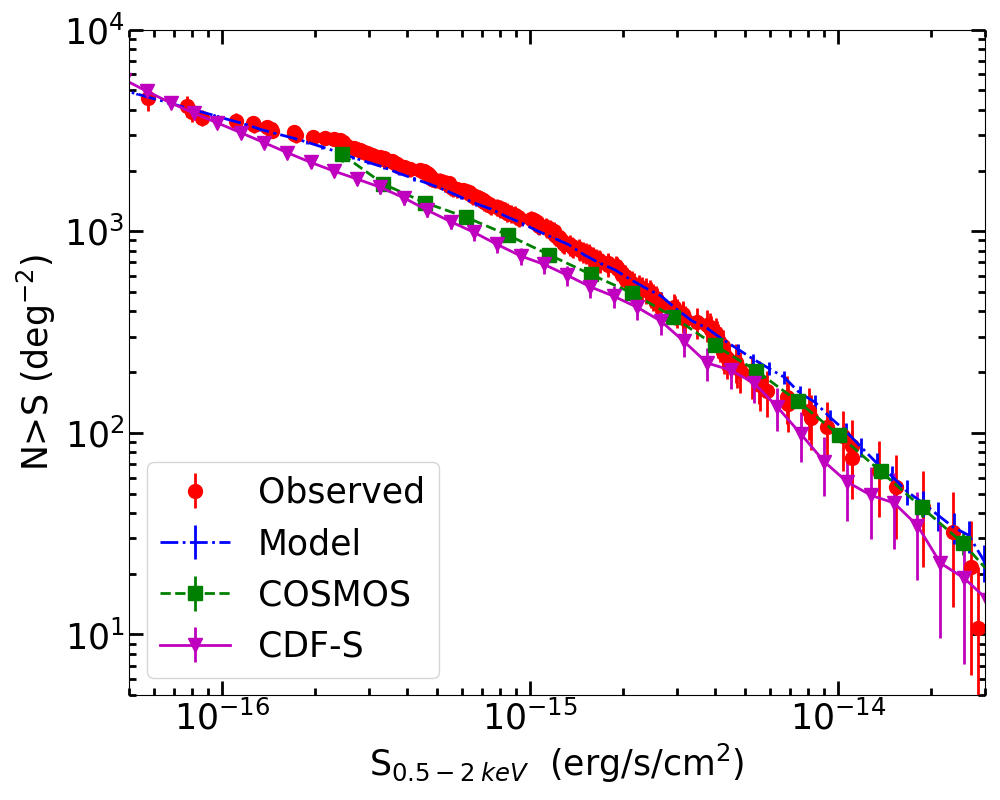}  \vspace{-0.05cm}\includegraphics[height=8cm, width=9cm, keepaspectratio]{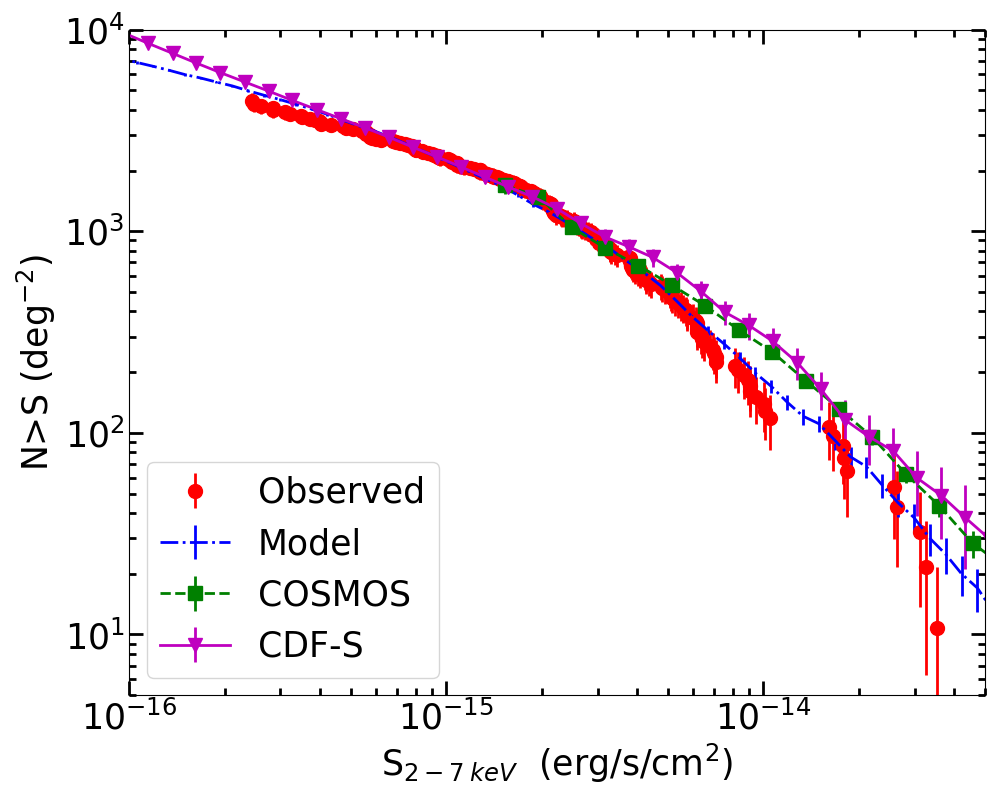}
\caption{The cumulative number counts (number of sources brighter than a given flux) for the main source catalog (red dots) in the full (top), soft (middle), and hard (bottom) bands. The blue dot-dashed line represents the cumulative number of sources from our mock catalog. For comparison, we plot the log(N)-log(S) relations found in the 7Ms \textit{Chandra} Deep Field-South by \citet[][magenta line]{Luo17}, and in the COSMOS field by \citet[][green dotted line]{Civ16}.}
\label{fig:logNlogS}
\end{figure}

Finally, we computed the cumulative number of sources, $N($>$S)$, brighter than a given flux ($S$) in each X-ray band. 
To this goal, we computed the sky-coverage (i.e., the sky area $\Omega$ covered as a function of the flux limit) of the J1030 field to correct the incompleteness of our catalog.
We computed our sky-coverage by dividing the number of output sources flagged as ``good'' in our simulations by the number of input sources as a function of input flux and then multiplying for the total geometric area of the J1030 field covered by \textit{Chandra}. The sky-coverage values were then fitted with a spline to obtain a smooth monotonically increasing function. The sky coverage in the full, soft, and hard bands are plotted in Fig. \ref{fig:sky_coverage}.
The derived flux limits over the central $\sim1$ arcmin$^2$ region are $\sim3\times10^{-16}$, $6\times10^{-17}$, and $2\times10^{-16}$ erg cm$^{-2}$ s$^{-1}$ in the full, soft, and hard bands, respectively, making the deep \textit{Chandra} survey in the SDSS J1030+0524 field the fifth deepest X-ray survey field achieved so far (see Fig. \ref{fig:surveys} for area-flux curve comparison with other surveys).
Once the sky coverage is known, the cumulative source number was computed using the equation:
\begin{equation}
 N(>S)=\sum_{i=1}^{N_S}\frac{1}{\Omega_i} \quad deg^{-2}
\end{equation}
where $N_S$ is the total number of detected sources in the field with fluxes higher than $S$, and $\Omega_i$ is the sky coverage associated with the flux of the i-th source. 
%The variance of the source number counts is therefore defined as
%\begin{equation}
% \sigma_i^2=\sum_{i=1}^{N_S}\Bigg(\frac{1}{\Omega_i}\Bigg)^2
%\end{equation}
The three log(N)-log(S) relations in the three X-ray bands are reported in Fig. \ref{fig:logNlogS}. The red points represent the cumulative number of sources of our J1030 field, while the log(N)-log(S) of our mock catalog are shown as the blue dot-dashed line. For comparison, we also plotted the log(N)-log(S) relations found in the 7Ms (magenta line) \textit{Chandra} Deep Field-South by \citet{Luo17}, and the one (green solid line) found in the COSMOS field by \citet{Civ16}. 
From Fig.\ref{fig:logNlogS} we conclude that the log(N)-log(S) relations derived in the J1030 field are in general agreement with those from the literature. 
Besides cosmic variance, we caution that some of the differences among the log(N)-log(S) seen in Fig. \ref{fig:logNlogS} could be produced by systematic uncertainties in the different methods to derive the sky coverage and the individual source fluxes in the various surveys.

\section{Summary}\label{conclusions}

We have presented the X-ray source catalog for the deep \textit{Chandra} survey in the SDSS J1030+0524 field, centered on a region that shows the best
evidence to date of an overdensity around a $z\sim6$ and an overdensity of galaxies at $z=1.7$. 
This field has been observed with 10 \textit{Chandra} pointings for a total exposure time of $\sim$479 ks and covers an area of 335 arcmin$^2$.
Furthermore, the J1030 field is part of the Multiwavelength Yale-Chile survey, and has been entirely observed by \textit{Spitzer} IRAC, LBT/LBC ($r$, and $z$ bands), and CHFT/WIRCam ($J$-band), making J1030 a legacy field for the study of large scale structures around distant accreting SMBHs.
Our main results are the following:
\begin{itemize}
\item The \textit{Chandra} source catalog contains 256 X-ray sources that were detected in at least one X-ray band (full, soft, and hard) by \textit{wavdetect} with a threshold of $10^{-4}$, and filtered by AE with a binomial probability threshold of $2\times10^{-4}$. We assess the binomial probability threshold by producing three X-ray simulations that mirror our \textit{Chandra} observation, obtaining a completeness of 95\% (full band), 97\% (soft band), and 91\% (hard band), while the reliability levels are 95\% (full band), 96\% (soft band), and 95\% (hard band).
\item We have achieved X-ray flux limits over the central $\sim1$ arcmin$^2$ region of $\sim3\times10^{-16}$, $6\times10^{-17}$, and $2\times10^{-16}$ erg cm$^{-2}$ s$^{-1}$in the full, soft, and hard bands, respectively, making the J1030 \textit{Chandra} field the fifth deepest X-ray survey in existence, after the CDF-S and the CDF-N surveys, the AEGIS-X survey, and the SSA22 survey. 
\item Based on the multi-band observations of this field, including $r$ and $z$ band data from LBT/LBC, $J$-band imaging from the CFHT/WIRCam, and $4.5\, \mu m$ from \textit{Spitzer} IRAC, we used a likelihood ratio analysis to associate optical/IR counterparts for 252 (98.4\%) of the 256 X-ray sources, with an estimated 95\% reliability. 
\item Finally, we computed the cumulative number of sources in each X-ray band finding that it is in general agreement with both our simulations and those from the CDF-S, the CDF-N, and COSMOS fields.
\end{itemize}

\begin{acknowledgements}
The scientific results reported in this article are based on observations made by the Chandra X-ray Observatory.
We acknowledge the referee for a prompt and constructive report.
We acknowledge financial contribution from the agreement ASI-INAF n. 2017-14-H.O.
We thank P. Broos for providing great support for the analysis of our simulations with AE, and H. M. G\"{u}nther for the support provided for using MARX.
We also thank B. Luo for providing us the log(N)-log(S) of the 7Ms CDF-S.
FV acknowledges financial support from CONICYT and CASSACA through the Fourth call for tenders of the CAS-CONICYT Fund, and CONICYT grants Basal-CATA AFB-170002.
DM and MA acknowledge support by grant number NNX16AN49G issued through the NASA Astrophysics Data Analysis Program (ADAP). Further support was provided by the Faculty Research Fund (FRF) of Tufts University.
 \end{acknowledgements}

\appendix

\section{Likelihood-ratio method}\label{appendix}

As described in \S \ref{completeness_reliability}, after producing simulations that mirror our \textit{Chandra} observation of J1030 and using \textit{wavdetect} to detect the sources on these simulated fields, we needed a numerical method to disentangle output sources that actually correspond to input ones from those that are spurious detections. To this purpose, we used a likelihood-ratio (LR) method to match output with input sources. The LR method we adopted was already used in past works to match sources detected at different wavelengths (e.g., \citealt{SutSan92}; \citealt{Cili03}; \citealt{Brusa07}; \citealt{Luo10}), and is available at: \url{https://github.com/alessandropeca/LYR_PythonLikelihoodRatio}. For an input simulated candidate with a flux $f$ at an angular separation $r$ from a given X-ray output, the LR is defined as in Equation \ref{eq:ciliegi2003}:
\begin{equation}
 LR = \frac{F(r)q(f)}{n(f)}
\end{equation}

In Equation \ref{eq:ciliegi2003} we assumed that $F(r)$ (the probability distribution function of the angular separation) follows a Gaussian distribution (e.g., \citealt{Zam99}):
\begin{equation}
F(r)= \frac{1}{2\pi \sigma_{pos}^2}exp\Bigg(\frac{-r^2}{2\sigma_{pos}^2}\Bigg)
\end{equation}
where $\sigma_{pos}$ is the 1$\sigma$ positional error of the X-ray detected sources computed as $\sigma_{pos} = PSF_{Radius}/\sqrt{C}$ (\citealt{Pucc09}), $C$ are the net, background-subtracted, counts computed by AE, and $PSF_{Radius}$ is evaluated with the estimate at the 90\% encircled energy radius (at E = 1.4 keV) at off-axis ($\theta$) as $PSF_{Radius}=1"+10"(\theta/10')^2$ (\citealt{Hick06}).

The flux-dependent surface density of the background sources, $n(f)$, is estimated using our sample of input simulated sources that are at an angular separation inside an annulus from any of the output detected sources ($r_{in}=5"$ and $r_{out}=30"$; e.g., \citealt{Luo10}). Input sources that fall inside the annular regions are considered as background sources.

$q(f)$ is the expected flux distribution of the real counterparts, and is not directly observable. To derive an estimate of $q(f)$, the LR method selects all input sources within $r_{in}=2"$ from any detected source. The flux distribution of these sources is denoted as $total(f)$, which was then background-subtracted to derive:
\begin{equation}\label{real}
real(f) = total(f) - \pi r_{in}^2 N_{out} n(f)
\end{equation}
where $N_{out}$ is the total number of X-ray detected sources. 

\begin{figure}
 \centering
  \vspace{-0.2cm}\includegraphics[height=8cm, width=8cm, keepaspectratio]{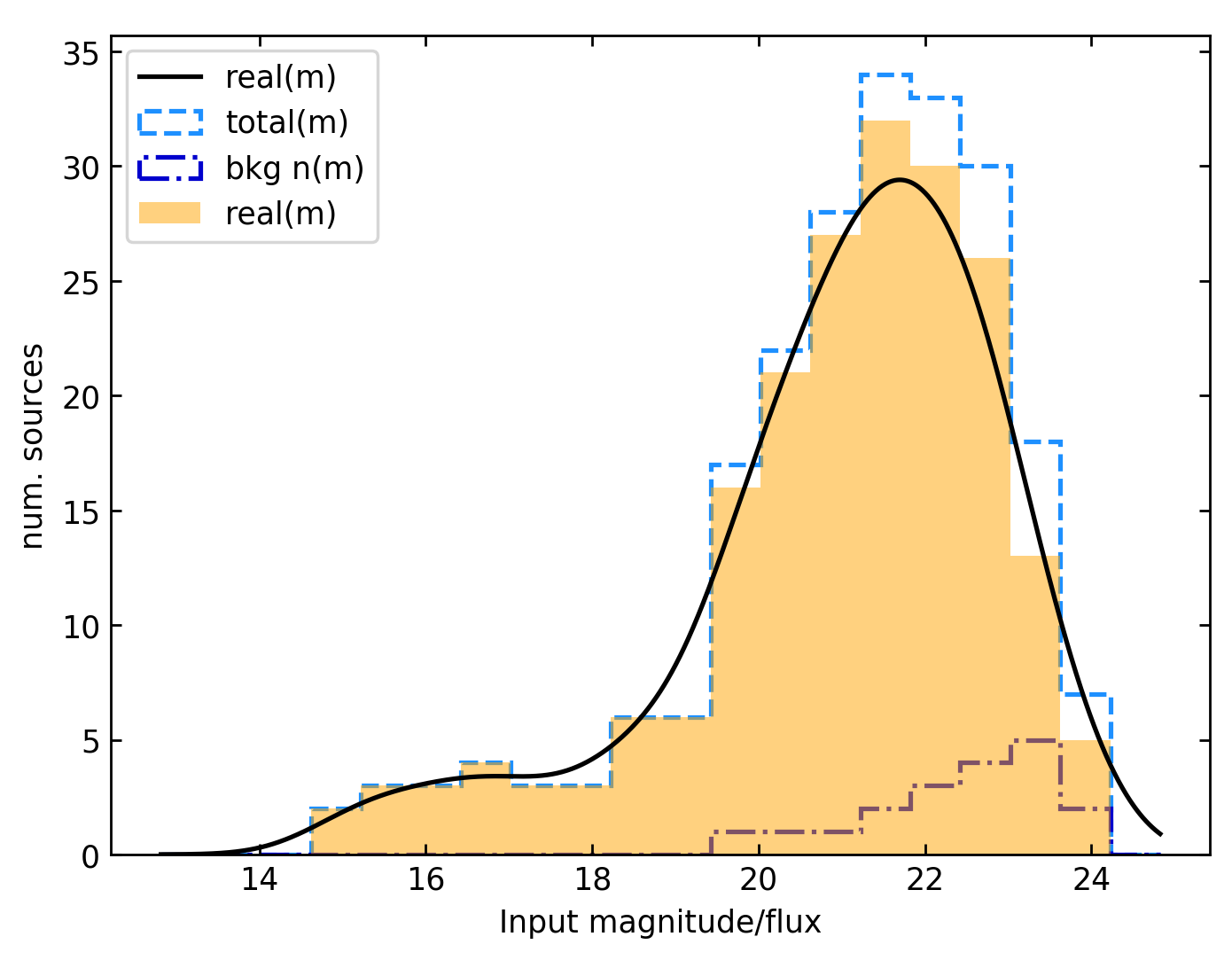}
\caption{Magnitude distribution of the $J$-band counterparts ($q(f)$, black line) and the magnitude distribution of the background sources ($n(f)$, blue dot-dashed line). The cyan dashed line represents the $total(f)$ (see \S \ref{appendix}) while the orange shaded area is the $total(f)$ defined as in eq. \ref{real}.}
\label{fig:mag_counter}
\end{figure}

An example of $real(f)$, $total(f)$, $n(f)$, and $q(f)$ distribution for the $J$-band counterparts is reported in Fig. \ref{fig:mag_counter}.
Due to the magnitude limits of the input catalog, we were only able to detect a fraction $Q$ of all the true counterparts (see \S \ref{completeness_reliability} for the definition of $Q$). Thus the expected flux distribution of the counterparts $q(f)$ is derived by normalizing $real(f)$ and then multiplying by $Q$:
\begin{equation}
q(f)= \frac{real(f)}{\sum_i real(f)_i}Q
\end{equation}

Having computed the values of $q(f)$, $n(f)$, and $f(r)$, our LR method calculates LR values for all the input sources within $r_{LR}=5"$ from each output detected source.
%Postage-stamp images for the X-ray catalog sources are reported in Fig. \ref{fig:optical_counterpart} The images are color composites of the LBT/LBC r, z, and CFHT/WIRCam J bands. In Fig. \ref{fig:irac_counterpart} we show 17 X-ray sources that, among the four optical/IR band explored, have a counterpart only at 4.5 $\mu m$.

\bibliographystyle{aa}
\bibliography{riccardo}

\end{document}